\documentstyle[aps,prb,multicol,epsf]{revtex}
\def\beq{\begin{equation}}
\def\eeq{\end{equation}}
\begin{document}                
\title{History-Dependence and Ageing in a Periodic Long-Range
Josephson Array}
\author{P. Chandra,$^1$ M.V. Feigelman,$^2$ L.B. Ioffe$^{2,3}$ and
D.M. Kagan$^2$}
\address{$^1$NEC Research Institute, 4 Independence Way, Princeton NJ
08540}
\address{$^2$Landau Institute for Theoretical Physics, Moscow, RUSSIA}
\address{$^3$Department of Physics, Rutgers University, Piscataway, NJ 08855}

\maketitle
\begin{abstract}
History-dependence and ageing are studied in the low-temperature
glass phase of a long-range periodic Josephson array.
This model is characterized by two parameters, the number of wires
($2N$) and the flux per unit strip ($\alpha$); in the limit $N
\rightarrow \infty$ and fixed $\alpha \ll 1$ the dynamics of the model
are described by the set of coupled integral equations, which
coincide with those  for the $p=4$ {\em disordered} spherical model.
Below the glass transition we have solved these equations numerically
in a number of different regimes.

We observe power-law ageing after a fast quench with an exponent that
decreases rapidly with temperature.
After slow cooling to a not-too-low temperature, we see ageing
characterized by the appearance of a new time scale which has a power
law dependence on the cooling rate.
By contrast if the array is cooled slowly to very low temperatures the
ageing disappears.
The physical consequences
of these results in different cooling regimes are discussed
for future experiment.

We also study the structure of the phase space in the low temperature
glassy regime.  Analytically we expect an exponential
number of metastable states just below the glass transition
temperature
with vanishing mutual overlap, and numerical results indicate that this 
scenario remains valid down to zero temperature.  Thus in this
array there is no further subdivision
of metastable states. We also investigate the probability to evolve to
different states given a starting overlap, and our results suggest
a broad distribution of barriers.
\end{abstract}


\section{Introduction}

The essential features underlying the physics of glass formation remain
elusive.  A glassy system has a ``memory'' of its past history; it
breaks ergodicity without thermodynamic selection of a unique state
thereby defying description by the standard Gibbs methods.  Though
glass formation in the absence of disorder is a widespread phenomenon,
it remains poorly understood even at the mean-field level.  Several
infinite-range models for disorder-free glassiness have been
proposed;\cite{activity} the
majority were studied via a mapping to intrinsically random
systems.\cite{disorder}
This approach relies on the observation that certain infinite-range
periodic and random
models have identical leading order (in $1/N$) high-temperature
expansions.  This correspondance is assumed to persist at
low temperatures.  Crystalline phase(s), which could appear
in a periodic model, are missed by such a treatment; however
the usual working assumption is that they are dynamically inaccessible.
Recently a glass transition in a periodic, long-range Josephson
array has been identified and characterized\cite{Chandra95,Chandra96a}
using a generalized
Thouless-Anderson-Palmer (TAP) approach.\cite{Thouless77}
This model can be studied directly (without a mapping to a disordered
system) because it has {\sl two} expansion parameters: the first
is the usual $1/N$ inherent to a long-range model; the second
is $\alpha$, the flux per unit strip, where the condition
$\alpha \ll 1$ further restricts the number of contributing
diagrams
to be a small subset of all those appearing
to $O(\frac{1}{N})$.
The existence of $\alpha$ as an expansion parameter suggests
that the crystalline phase(s) are inaccessible (or do not exist)
for small values of $\alpha$. We note that for $\alpha = 1$
there exist non-trivial ground-states for an array with prime $N$
which are certainly missed by perturbative methods.
In this particular case, any sequence of complex numbers (representing
the order parameters associated with each wire)
of unit magnitude whose discrete Fourier transform has a flat
spectrum yields a ground-state.\cite{Stephens96} This mathematical problem has been
studied extensively in the context of number theory,
beginning with the work of Gauss.\cite{Schroeder86}

Another key feature of the periodic
Josephson array is that it can be
constructed in the laboratory;
the physical requirements imposed
by the theory on its realization
have been
discussed elsewhere.\cite{Chandra96b}
Measurement of the a.c. response
to a small time-varying
field in the
fabricated array should probe its diverging
relaxation time at the glass transition.\cite{Chandra96a,Chandra96b}
In this paper we study the behavior of this
periodic network in its low-temperature non-ergodic regime.
We find both history-dependence
and ageing, and the physical consequences of these  results
for different cooling regimes are discussed for future experiment.

\centerline{\epsfxsize=8cm \epsfbox{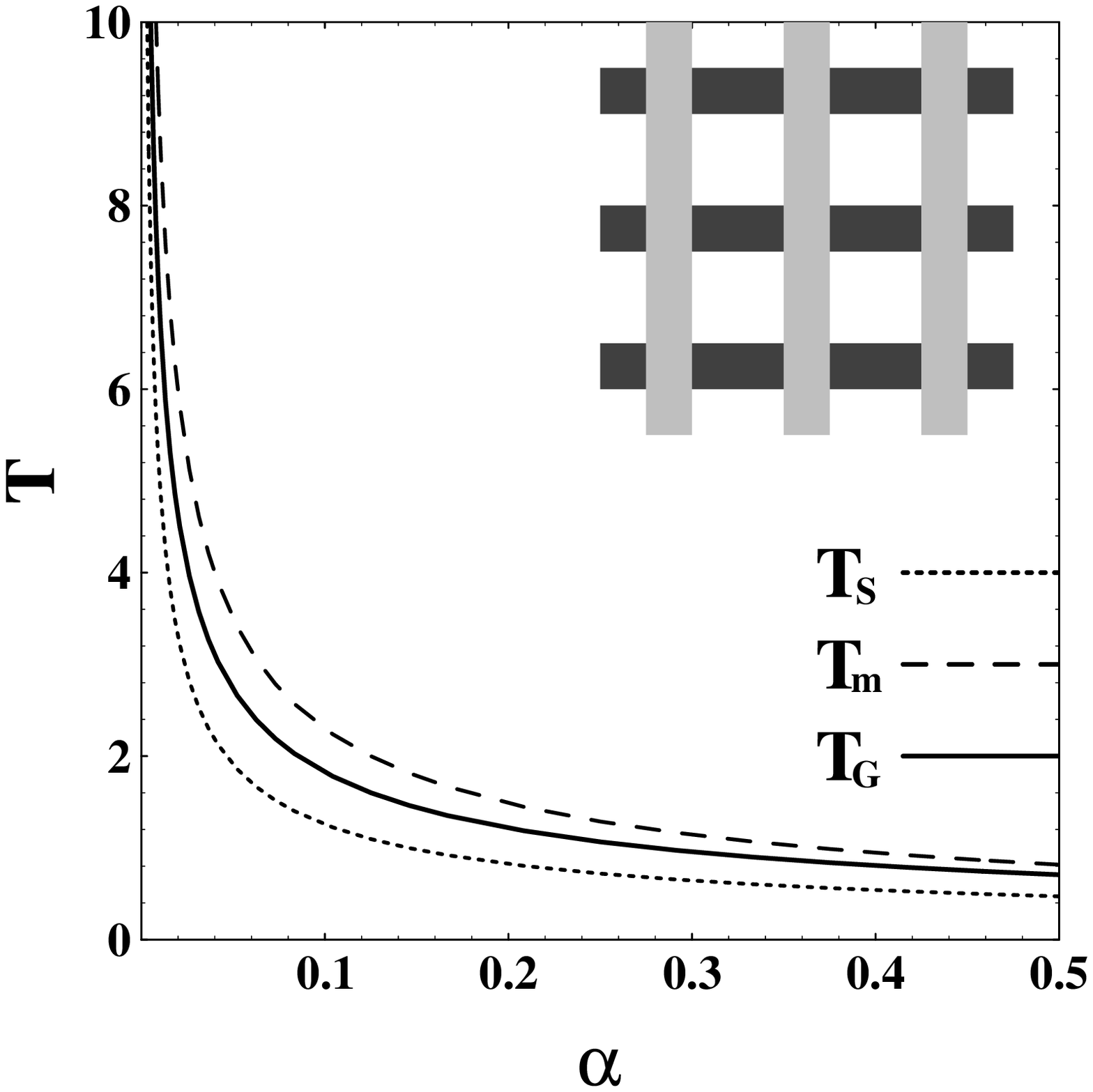}} 
{\footnotesize {\bf Fig~1. The phase diagram of the array (inset)}. Here
$T_G$, $T_m$ and $T_S$ are the temperatures associated with the dynamical
instability, ``superheating'', and a speculated equilibrium transition
as discussed in the text.
}\vspace{0.5in} 

The system of our study is a stack of two mutually perpendicular sets of
$N$ parallel thin superconducting
wires with Josephson junctions at each node (Figure 1)
that is placed in an external tranverse field $H$.  The classical thermodynamic
variables of this array are the $2N$ superconducting phases associated with
each wire.  In the absence of an external field the phase differences
would be zero at each junction, but this is not possible for finite
$H$ and the phases are thus frustrated.  Here we assume that the
Josephson couplings are sufficiently small so that the induced fields
are negligible in comparison with $H$, an important point for the
experimental realization of this network.\cite{Chandra96b}
We can therefore describe
the array by the Hamiltonian
\beq
{\cal H} = - \sum_{m,n}^{2N} s_m^{*} {\cal J}_{mn} s_n
\label{H}
\eeq
where ${\cal J}_{mn}$ is the coupling matrix
\beq
\hat{\cal  J} = \left( \begin{array}{cc}
0 & \hat{J} \\
\hat{J}^\dagger & 0
\end{array}
\right)
\label{J}
\eeq
with $J_{jk} = \frac{J_0}{\sqrt{N}} \exp(2\pi i \alpha jk /N)$ and $1 \! \leq
\! (j,k) \! \leq \! N$ where $j(k)$ is the index of the horizontal (vertical)
wires;
$s_m = e^{i\phi_m}$  where the $\phi_m$ are the superconducting phases of the
$2N$ wires.
Here we have introduced the flux per unit strip, $\alpha = NHl^2/\Phi_0$,
where $l$ is the inter-node spacing and $\Phi_0$ is the flux quantum;
the normalization has been chosen so that $T_G$ does not scale with
$N$.

Because every horizontal (vertical) wire is linked to every vertical
(horizontal) wire, the connectivity in this model is high ($N$) and
it is accessible to a mean-field treatment.\cite{Vinokur87}  For ${1\over N} \ll \alpha < 1$
there exist an extensive number of metastable solutions separated
by barriers that scale\cite{Chandra95}  with $N$.
A similar long-range network with
disorder was previously found to display a spin glass
transition\cite{Vinokur87} for
$\alpha \gg {1\over N}$; in the absence of short-range phase
coherence between wires $(\alpha \gg 1$), it was mapped onto the
Sherrington-Kirkpatrick model.\cite{Sherrington75}
  Physically this glassy behavior
occurs because the phase differences associated with the couplings, $J_{jk}$,
acquire random values and fill the interval $(0,2\pi)$ uniformly.
For the periodic case, this condition is satisfied in the ``incommensurate
window'' ${1\over N} \ll \alpha \le 1$; here the magnetic unit cell
is larger than the system size so that the simple ``crystalline'' phase is
inaccessible.\cite{Chandra95} There are thus no special field values where the
number of low-temperature solutions are not extensive, in contrast
to the situation for $\alpha  > 1$. Of course, we cannot exclude the
possibility that at low temperatures a single low-lying state appears;
however it seems very likely that even in this case it will be
irrelevant because the system is trapped in one of the many
metastable glassy states. Furthermore we have not observed such states
in our direct numerical simulations of $\alpha=1$ array.

In the thermodynamic limit of $N \rightarrow \infty$ (with fixed area),
the system is accessible to a static and dynamical high-temperature
study;\cite{Chandra95,Chandra96a}
this was performed using a modified TAP method.\cite{Thouless77}  This will
be discussed in more detail when the coupled dynamical equations are
derived in Section II; at temperatures above the dynamic instability
the system is still in equilibrium and these equations can be
solved analytically.  The glass transition is characterized by
a diverging relaxation time and a jump in the Edwards-Anderson
order parameter, and does {\sl not} coincide with an
accompanying static transition. This calculated glass transition
temperature ($T_G$) assumes that cooling from high temperatures occurs
infinitesmally slowly; in practice, the system's
effective glass transition
temperature ($T_G^{eff} > T_G$) will correspond to
the system's failure to equilibrate and will be a function of
its finite cooling time-scale. For the sake of notational simplicity,
we will not distinguish between the effective and the ``adiabatic''
glass transition temperature in the text that follows unless
explicitly
stated.

Below the glass transition temperature ($T_G$),
the system can no longer equilibrate.
In contrast to the situation for $T > T_G$,
the
response and the correlation functions
are {\sl not} a function of time
differences but rather are history-dependent;
furthermore we can no longer relate them
with the fluctuation-dissipation theorem.
This makes the solution of these equations
rather complicated.  The low-temperature
state of the system is determined by the
sample history; here we shall consider
the simplest scenario in which the
temperature is reduced linearly from
an initial $T_i$ slightly above $T_G$
to a final temperature $T_f (< T_G$)
in a time $t_c$.  The final state of
the system at time $t > t_c$ is controlled
by both time-scales $t$ and $t_c$, and
we expect qualitatively different behavior
depending on their relative magnitude.
Here we shall focus on the limiting cases
$t \gg t_c$ and $t = t_c$.

Physically in the limit of very fast cooling ($t \gg t_c$),
we expect the state immediately following the quench to have
minimal overlap with that observed at time $t$ so that
$D_{tt'\rightarrow 0} = 0$.  Assuming that the solutions
of the dynamical equations are well-defined in the limit
when the microscopic time-scale goes to zero, we have only
two time-scales in the problem: $t$ and $t'$.  In this
case $D_{tt'}$ may depend only on their ratio $t'/t$ on dimensional
grounds:
$D_{tt'} = d({t'\over t})$.  Since $d(0) = 0$
we expect
$\lim_{t'/t \rightarrow 0} d({t'\over t}) = ({t'\over
t})^\gamma D_0$, and this functional
form of $D_{tt'}$ remains\cite{Cugliandolo93,Cugliandolo94}
at larger values of $\left(t'\over t\right)$.
Similarly in this fast-cooling regime we expect
$\lim_{t'/t\rightarrow 0} G_{tt'} = {1\over t'}
\left ({t'\over t}\right)^\gamma G_0$,
so that in the limit of small $(t'/t)$ there is
a quasi-FDT relation ($ G \propto D'$),
which remains valid\cite{Cugliandolo93,Cugliandolo94}
over the full range $0 < (t'/t) <  1$.

The other limiting case is that of slow cooling ($t = t_c$),
where we expect that the correlation and response functions
will have their only time-dependence through the specified
cooling process $\tau(t)$ where
$\tau \equiv \left( {T_G- T\over
T_G}\right)$:
$D_{tt'} = d [\tau (t), \tau(t')]$ and
$G_{tt'} = {d\tau(t')\over dt'} g[\tau (t), \tau(t')]$.
If this conjecture is correct, it implies that the
system has a non-decaying memory of its past history.
Of course this assumes again that the solutions to the
dynamical equations are well-defined in the limit of vanishing
microscopic time-scales.  For the Sherrington-Kirkpatrick
model,\cite{Sherrington75} this ``adiabatic ansatz'' for linear $\tau (t)$
leads\cite{Ioffe88} to the solutions $d(t,t') = \tau_{t'}$  and $g(t,t') =
2\tau_{t'}$, which result in a field-cooled susceptibility in
qualitative agreement with a previous equilibrium result.\cite{Sommers85}
However application of the same method to the $p=4$ Ising
model leads to a delta-function solution for the
response function;\cite{Feigelman95}
this form is worrisome given the original adiabatic nature of
the ansatz.  However, in principle the response could be broadened
to be long compared to the microscopic time-scale but short compared
to the final measurement time.\cite{Feigelman95}
When we started our numerical study of this problem, our expectations were
therefore that the only time scale in the slow cooling process is set by the
cooling rate, $t^* \equiv ({d\tau\over dt})^{-1}$, and that the response
and the correlation functions depend only on the ratios $t/t^*$ and $t/t^*$ as
indicated in the $t=t_c$ column of Table 1. This hypothesis turned
out to be only partially correct. Instead we have found that indeed just below
$T_G$ the correlation and response functions depend only on the ratios $t/t^*$
and $t/t^*$ but $t^*$ is a {\em new} time-scale
$t^* \equiv \left(dt\over d\tau\right)^{1 - \eta(T)}$
that emerges in this cooling regime; however there is evidence
that the exponent $\eta$ vanishes at low temperatures,
indicating that ``adiabatic cooling'' is recovered.
We note that a finite value of $\eta$ implies a decaying
memory and hence ageing, and thus the disappearance of $\eta$ means that the
system recovers its memory and thus ``rejuvenates''
at low temperatures.

Our expectations for the three cooling regimes are summarized
in Table I. In Section II. we derive the dynamical equations
for the array in the low-field $(\alpha \ll 1)$ regime.
The system goes out of equilibrium at a temperature $T_G$
(determined by the cooling rate), and we study its low-temperature ($T < T_G$)
behavior numerically in all three cooling regimes; our results are presented
in Section III. 

We also probe the structure of phase space in this glassy regime (Section
IIID) where the periodic array is stuck in one of an extensive number of
metastable states. \cite{Chandra95} 
What happens to these states when the temperature is
decreased?
Each state, formed at $T_G$, may evolve smoothly; alternatively
it can further subdivide as in the case of the (disordered)
Sherrington-Kirkpatrick model.
We find that in this periodic array there is no further subdivision
down to zero temperatures, and that the number of metastable states
remains exponentially large.  We also learn that though
these states are equidistant in phase space, their associated
basins of attraction are rather complicated.
We conclude with a discussion of the physical consequences for experiment
(Section IV), a brief summary and with projects for future work.

\begin{table}[h]
\caption{Three distinct regimes and their associated
response ($D_{tt'}$ and correlation ($G_{tt'}$)
functions when the system is quenched out-of-equilibrium ($t <
t_R$);
here $t$, $t_r$ and $t_c$ are the measurement, relaxation and cooling
times
respectively, and $t^*$ is a time-scale
which is a function of the cooling rate (${d\tau}\over
{dt}$).}
\begin{tabular}{cccc} $t \gg t_c$
        &  $t  \gtrsim  t_c$
        &  $t = t_c$                     \\
\tableline 
$D_{tt'} = \left( {t' \over t} \right)^\gamma D_0$
& ???
& $D_{tt'} = d \left( {t' \over t^*},{t \over t^*} \right) $ \\
$ G_{tt'} = x D'_{tt'}$
&  ???
& $G_{tt'} = {1\over t^*} g\left( {t' \over t^*},{t \over t^* } \right)$ \\
& & $t^* = f\left( {dt\over d\tau} \right)$
\end{tabular}
\end{table}

\section{The Dynamical Equations}

The time-dependent properties of a glass
are particularly distinctive, and thus we begin our study
with a discussion of the self-consistent equations
describing the dynamics of the periodic Josephson array.
For the sake of convenience, we shall refer to the network
model
described by the Hamiltonian
(\ref{H}) in ``magnetic'' language so that each
``spin'' $s_m = e^{i\phi_m}$ is associated with the phase
of a
superconducting wire labelled by ``site index'' $m$.
We presume that the system's long-time behavior can
be described by that of soft spins with pure relaxational
modes, since intrinsic single-spin dynamics
should be irrelevant on these time-scales. More specifically we introduce
a ``constraining potential'' $V(s_i) = V_0(|s|^2 - 1)^2$
at each site that restricts the magnitude of
each spin, $|s_i| \approx 1$, and then assume the equations
of motion
\beq
\tau_b \dot{s_i} = -\frac{1}{T} \frac{\partial ({\cal H} + V)}{\partial
s_i^*} + \zeta_i
\label{langevin}
\eeq
where $\tau_b$ is a microscopic time-scale; the effects of a
coupled
heat bath are represented by a time-varying Gaussian random
field $\zeta_i(t)$ with zero mean and variance
\beq
\langle \zeta_i(t)  \zeta_j (t')\rangle = 2\tau_b \delta_{ij}\delta(t
- t')
\label{noise}
\eeq
where the brackets refer to an average over the noise.
This Langevin equation (\ref{langevin})
reproduces the dynamics of the overdamped
Josephson array\cite{Chandra96b} with effective
resistance $R$ if $\tau_b= \frac{R_Q}{R}
\frac{\hbar}{T}$
and $V_0 \rightarrow \infty$ where $R_Q = \frac{\hbar}{4e^2}$;
here
$R = \frac{R_0}{N}$ where $R_0$
is the individual
junction resistance since the circuit elements in the long-range
array operate in parallel.\cite{Chandra96b}

The physical quantities of interest here are the dynamical
cumulants
of the ``spin'' variables
averaged over the noise.  In particular, we would
like to determine the time-dependant two-spin
correlation function
\beq
D_{mn}(t,t') = \langle s_m(t) s_n^*(t') \rangle
\label{D}
\eeq
and the linear response
\beq
G_{mn}(t,t') =    \frac{\partial\langle
s_m(t)\rangle}{\partial h_n(t')}, \qquad t > t'
\label{G}
\eeq
where again the angular brackets refer to an average over
$\zeta_i(t)$.  We note that $G_{mn} (t,t')$ is a susceptibility with
respect to a ``theorist's'' field $h_n$ conjugate to a spin $s_n =
e^{i\phi_n}$,
and thus cannot be measured directly; however once known,
$G_{mn}(t,t')$
can be used to determine the ac-response to a time-varying physical
external field $H(t)$ which is experimentally
accessible.\cite{Chandra96a}

Exploiting the fact that the Langevin noise $\zeta_i(t)$ is Gaussian
and uncorrelated in space and time, we can perform the noise averages
in (\ref{D}) and (\ref{G}) using a functional integral
formulation;\cite{Dominicis76,Janssen76}
details of this technique can be found elsewhere.\cite{Sompolinsky82}
For example, the two-spin dynamical correlation
function is given by the expression
\beq
D_{mn}(t,t') = \langle s_m(t) s_n^*(t')\rangle
= \int s_m(t) s_n^*(t) \exp
{\cal A}[s,\hat{s}]  {\cal D}s {\cal D}\hat{s}.
\label{Dav}
\eeq
with the action
\beq
{\cal A}[s,\hat{s}] = \int dt \left[ \hat{s} \left(\tau_b\dot{s} + \frac{1}{T}
\frac
{\partial ({\cal H} + V)}{\partial s^*} \right) +
 \tau_b\hat{s}^2 + h.c. \right]
\label{A}
\eeq
where we have neglected the term arising from the functional
Jacobian since it does not affect the system's long-time behavior
provided
that causality is maintained; the latter is ensured by the
diagrammatic treatment described below.
The auxiliary field $\hat{s}$ in a correlation function\cite{Martin81}
acts like a response term $\frac{\partial}{\partial h(t)}$ so that
$G_{m n}(t,t') = \langle s_m(t)\hat{s}_n(t')\rangle$ can be determined
by an expression analogous to (\ref{Dav}).

\centerline{\epsfxsize=8cm \epsfbox{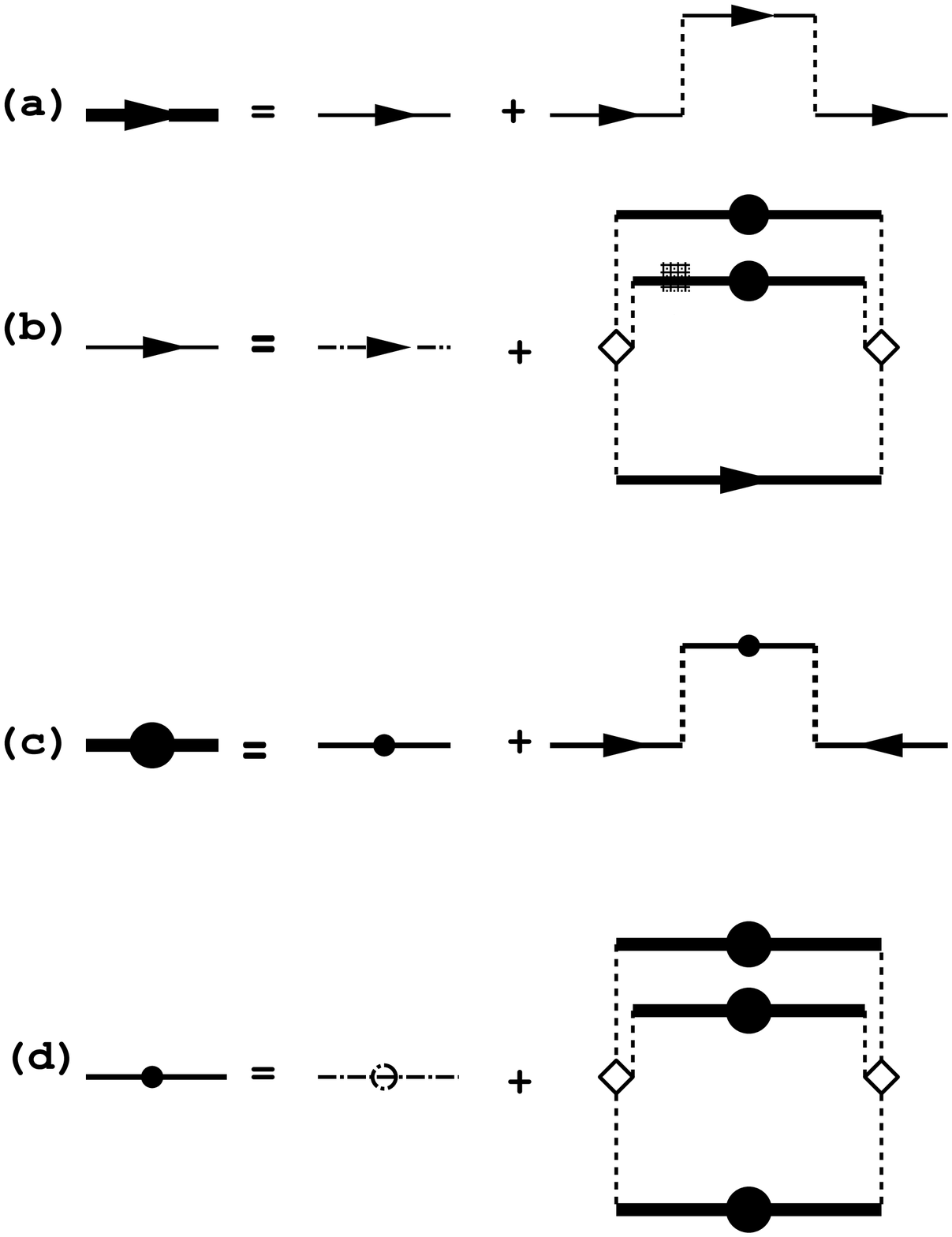}} 
{\footnotesize {\bf Fig 2. The diagrammatic expansion (to leading order in
$1/N$)  for the (a) response and the (b) correlation functions.} Here
the coupling matrix $\cal{J}$ is represented by dashed lines, $\Gamma$ is
the single-site four-spin vertex shown as a diamond,
$\tilde{G}_\omega$ ($\tilde{D}_\omega$) is 
the
single-site irreducible response (correlation) function, 
$\bf G$ ($\bf D$) is
the full response
(correlation)
function, and $\tilde G_b$ ($\tilde D_b$) is the single-site bare
response (correlation) function.  The leading self-energy
contributions 
to $\tilde{G}_\omega$ and $\tilde{D}_\omega$
(in $\alpha \ll 1$) are displayed in (c) and (d) respectively.
}\vspace{0.5in} 

More specifically, we calculate the time-dependent correlation
and response functions, $D_{mn}(t,t')$ and $G_{mn}(t,t')$
respectively,
by resumming the terms in the $\frac{\cal H}{T}$ expansion of
the generating functional associated with
(\ref{A}) which are leading order in $\frac{1}{N}$;
this is a dynamical generalization\cite{Chandra95,Chandra96b}
of the Thouless-Anderson-Palmer
approach\cite{Thouless77} where feedback
through single-site Onsager terms play a crucial
role.
It is convenient to
determine these response and correlation
functions initially
in the frequency-domain.
The leading diagrams (in $\frac{1}{N}$)
for $\hat{G}_{\omega}$ are shown in Figure 2a; in the
thermodynamic limit ($\lim N
\rightarrow \infty$ with fixed array area) the subleading
corrections to the response can be safely neglected.\cite{Chandra96b}
Summing the
geometric series shown in Figure 2a, we obtain
\beq
\bf{G}_\omega =
\frac{1} {\tilde{G}^{-1}_\omega - \beta^2 (\bf{J J^{\dagger}})\tilde{G}_\omega}
\label{Gg}
\eeq
for the response function connecting wires of the same type
(horizontal/vertical); here we consider the local Green's
functions $\tilde{G}_{\omega}$ that is irreducible with
respect to the $J_{ij}$ lines.
The matrix $(J J^\dagger)_{ij}$ depends  only on the ``distance'' $i-j$
and acquires a simple form in Fourier space $(J J^\dagger)_p =
(J_0^2/\alpha) \theta(\alpha \pi - |p|)$; therefore in this
representation
the low-frequency contribution to the response
function\cite{Chandra96a}
is
\beq
G^{(0)} = \int {dp\over (2\pi)} \frac{\theta(\alpha \pi - |p|)}
{\tilde{G}^{-1}_\omega -
        \frac{(\beta J_0)^2}{\alpha} \tilde{G}_\omega} +
\frac{\theta(|p| - \alpha \pi)}{\tilde{G}_\omega^{-1}}
\label{GG}
\eeq
which is of primary interest in this problem.
We note that the static limit
$(\omega = 0$) of $T\tilde{G}^{-1}_\omega$ coincides
with the locator, $A(T)$, discussed in previous work
on this array;\cite{Chandra95} in the absence of
feedback terms $\tilde{G_0}^{-1} = 1$.  We also see
in (\ref{GG}) that there is a static instability
at $\tilde{G}_0^{-1} = G_c = \frac{\beta J_0}{\sqrt{\alpha}}$
in the vicinity of $T_0 = \frac{J_0}{\sqrt{\alpha}}$.  For
$\Theta = \frac{(T - T_0)}{T_0} \gg \sqrt{\alpha}$ all feedback
terms are negligible and $\tilde{G}^{-1}_\omega \equiv
\tilde{G_b}^{-1}(\omega)
= \beta A(T) - i\omega \tau_b$ where $\tau_b$ is a microscopic
time-scale.
However at lower temperatures, $\Theta \le \sqrt{\alpha}$, feedback
effects introduce an additional frequency-dependent part of
the local response
\beq
\tilde{G}_\omega = \tilde{G}_b(\omega) + \Sigma_\omega
\label{Glocal}
\eeq
as a self-energy $(\Sigma_{\omega}$ such that $\Sigma_0 = 0$) since we have
chosen our normalization so that $\tilde{G}_0^{-1} = \beta A(T)$.
The leading contribution in $\alpha$ to the self-energy
$\Sigma_\omega$
is displayed in Figure 2b. We note that the self-energy
contains a term $\bf\hat{G} = \beta_0 J^{\dagger} G J$ where ${\bf G}$ is the {\sl full}
response function and $\beta_0 = \frac{1}{T_0}$,
and it is therefore convenient to write
an equation for $\bf\hat{G}$.
We shall use the identity
\beq
{\bf G = (G - } \beta_0^2 {\bf J^{\dagger}  G J) + \hat G, \qquad\qquad
\hat G =} \beta_0^2 {\bf J^{\dagger} G J}
\label{Ghat}
\eeq
and note that, since the matrix $\bf J J^\dagger$ has only two eigenvalues,
$0$ and $T_0^2$,  the expression ${\bf (G - } \beta_0^2 {\bf J^{\dagger} G  J) }=
{\bf (1 - } \beta_0^2 {\bf J^{\dagger}  J) G} =
{\bf (1 - } \beta_0^2 {\bf J^{\dagger}  J)  \tilde{G}}$ projects only the zero eigenvalue contribution
of $\bf J J^\dagger$ in
the denominator
of (\ref{Gg}). We note that $\bf \tilde{G}$ is diagonal so that
the order of matrix multiplication is not important.
$\bf \hat G$ is equivalent
to the long-time part of the response function and is
proportional to $\alpha$ (cf. (\ref{GG}). Higher order
terms in the self-energy will thus contain higher powers
of $\alpha$, and thus we conjecture that for $\alpha \ll 1$
these terms can be neglected; this will be justified later
by the structure of the resulting solutions for $\bf G$ and $\bf D$.

The retardation of the self-energy terms (cf. Fig. 2b and 2d) significantly
affect the long-term behavior of the system. This
feedback results in a diverging relaxation time and a dynamical
instability\cite{Chandra96a} at a temperature $T_G = (1 + \Theta_G) T_0$
where $\Theta_G = -\frac{3^{3/4}}{2^{1/2}} \alpha^{1/4}$.

For temperatures $T > T_G$ the response and correlation functions
depend {\sl only} on the time-differences, and thus can be related
by the Fluctuation-Dissipation Theorem.  However for $T < T_G$ this
relation no longer holds and we must determine the correlation
function independently.  Analogous to our treatment
above, we introduce $\bf \tilde{D}$ and $\bf \hat{D}$ as in
equation (\ref{Ghat}):
\beq
{\bf D = (D - } \beta_0^2 {\bf J^\dagger D J) + \hat D, \qquad\qquad
\hat D =} \beta_0^2 {\bf J^\dagger D J}.
\label{Dhat}
\eeq
where again we use $\bf{ (D - } \beta_0^2 {\bf J^\dagger D J)}=
\bf{ (1 - } \beta_0^2 {\bf J^\dagger  J)\tilde{D}}$.
The leading diagrams (in $\frac{1}{N}$) for $\hat{D}_{\omega}$ are
shown in Figure 2c; the low-frequency contribution to
this correlation
function is
\beq
\hat{D}^{(0)} = \int {dp\over (2\pi)} \ \frac{\theta(\alpha \pi - |p|)} {\tilde{G}^{-1}_\omega -
        \frac{(\beta J_0)^2}{\alpha} \tilde{G}_\omega}
\hat{\Pi}_{\omega}  \ \frac{1}{\tilde{G}^{-1}_\omega -
        \frac{(\beta J_0)^2}{\alpha} \tilde{G}_\omega}
\label{Dd}
\eeq
where $\Pi$ is an on-site
self-energy contribution to $\tilde{D}_\omega$.
For long times and $\left( \frac{1}{N} < \right) \alpha \ll 1$
\beq
\hat{G}_{t_1,t_2}
=  \ \frac{\alpha}{a + 2\frac{\partial}{\partial t_1} \tau_b
- 2\Sigma_{t_1,t_2}}
\label{GGG}
\eeq
and
\beq
\hat{D}_{t_1,t_2} = \ \frac{\alpha}{a + 2\frac{\partial}{\partial t_1} \tau_b
- 2 \Sigma_{t1,t2}} \  \Pi_{t_1,t_2} \
\frac{1}{a + 2\frac{\partial}{\partial t_1} \tau_b
- 2 \Sigma_{t1,t2}}
\label{DD}
\eeq
where $a \equiv \beta (A - \frac{J_0^2}{\alpha A}) \approx 2 \Theta$
for $\Theta \gg \sqrt{\alpha}$ and $a \approx - \frac{\alpha}{\Theta}$
for $-\Theta \gg \sqrt{\alpha}$
and $A$ is the locator.\cite{Chandra96a}
Possible on-site self-energy terms, $\Sigma$ and $\Pi$,
displayed to leading order in $\alpha$  in Figures 2b and 2d, simplify in the
limit of $\alpha \ll 1$. In this regime, which
we consider here, we note from (\ref{GGG}) and
(\ref{DD})
that the slowly decaying parts of $\hat{G}$ and $\hat{D}$
scale with $\alpha$. Therefore the
dominant self-energy
contributions  contain the minimal number of these functions
(cf. Figs. 2c and 2d respectively).
We can therefore consider only the leading order feedback terms
\beq
\Sigma_{t1,t2} = \frac{3}{2}\hat{D}^2_{t1,t2} \hat{G}_{t1,t2}
\label{Sig}
\eeq
and
\beq
\Pi_{t1,t2} = \hat{D}^3_{t_1,t_2}  + 2\tau_b \delta_{t1,t2}
\label{Pi}
\eeq
where $\hat{G}$ and $\hat{D}$ are as in (\ref{GGG}) and (\ref{DD})
respectively, and we have approximated the four-spin vertex by
its static value $\Gamma = -1$.

We now have expressions for the correlation and the response functions
\begin{eqnarray}
&\left(a_{t_1} + 2 \tau_b {\frac{\partial}{\partial t_1}} - 2
\Sigma_{t_1,t_2}\right) \hat{G}_{t_1,t_2} &= \alpha \delta (t_1 - t_2)
\label{e1}\\
& \left(a_{t_1} + 2 \tau_b {\frac {\partial}{\partial t_1}} - 2
\Sigma_{t_1,t_2}\right) \hat{D}_{t_1,t_2} &= \int \hat{D}^3_{t_1,t_2}
\hat{G}_{t_2,t'} dt' + 2\tau_b \hat{G}_{t_1,t_2}
\label{e2}.
\end{eqnarray}
In order to solve these equations we also require a boundary
condition for $\hat {D}(t,t)$,
which
follows from the definition (\ref{Dhat}) and the condition that
$D_{i,i}(t,t) = 1$
so that
\beq
(1 - \alpha) \tilde{D}(t,t)  + \hat{D}(t,t)= 1.
\label{Dbound}
\eeq
We only need to consider equation (\ref{Dbound}) to order
$O(\alpha^{1/4})$ which allows us to neglect
self-energy corrections to $\tilde{D}$ (cf. Figure 2d) that
scale as $\alpha^{3/4}$
in this case $\tilde{D}(t,t) \approx \tilde{D}^{(0)}(t,t)$
then $\hat D(t,t) \approx 1 - D^{(0)}(t,t)$.
In the vicinity of the glass transition studied here, $-\Theta \sim
\alpha^{1/4} \gg \sqrt{\alpha}$, we have\cite{Chandra96a}
$\tilde{D}_{tt}^{(0)} = \tilde{G}_{\omega=0}^{(0)} \approx (1 +
\Theta)$
and therefore
\beq
\hat{D}_{t,t} = -\Theta_t.
\label{bc}
\eeq
This boundary condition
and equations (\ref{e1}) and (\ref{e2}) form a closed system of
equations.

We can simplify (\ref{e1}) and (\ref{e2})
by defining the parameters
\beq
t_0 = 2 \tau_b \alpha^{-3/4} \qquad\qquad \theta = - \Theta
\alpha^{-1/4}\qquad\qquad a_{t_1} = \tilde{a}_{t_1} \alpha^{3/4}\Theta_{t_1}
\label{scaling}
\eeq
and the rescaled functions
\begin{eqnarray}
&\hat{D}_{t_1,t_2}& = \alpha^{1/4} \theta_{t_1}
d\left(\frac{t_1}{t_0\theta_t},\frac{t_2}{t_0\theta_{t_2}}\right)
\label{dscal}\\
& \hat{G}_{t_1,t_2}& = \alpha^{1/4} \frac{1}{t_0}
g\left(\frac{t_1}{t_0\theta_t},\frac{t_2}{t_0\theta_{t_2}}\right)
\label{gscal}
\end{eqnarray}
to obtain the equations
\begin{eqnarray}
&\left(\tilde{a}_{t_1} + {\frac{\partial}{\partial t_1}}\right) 
g(t_1,t_2)  - 3 \theta^3_{t_1} \int^{t_2}_{t_1}
d^2(t_1,t')g(t_1,t')g(t',t_2) \theta_{t'} dt'& = \delta(t_1 - t_2)
\label{se1}\\
&\left(\tilde{a}_{t_1} + {\frac{\partial}{\partial t_1}}\right) 
d(t_1,t_2)  - 3 \theta^2_{t_1} \int^{t_1}_0
d^2(t_1,t')g(t_1,t')g(t',t_2) \theta^2_{t'} dt'& \nonumber \\
&\qquad\qquad- \theta_{t_1}^3 \int_0^{t_2} d^3(t_1,t') g(t_2,t') \Theta_{t'} dt' -
2g(t_1,t_2)& = 0
\label{se2}
\end{eqnarray}
where we have used the leading order (in $\alpha \ll 1$) self-energy
terms, (\ref{Sig}) and
(\ref{Pi}).
The boundary conditions, $d(t,t) = 1$ and
\beq
\left.g(t_1,t_2)\right\vert_{t_1 \rightarrow t_2^+} = 1 \qquad\qquad
\left.\frac{\partial}{\partial t_1} d(t_1,t_2) \right\vert_{t_1 \rightarrow t_2^-}
= 1
\label{bc2}
\eeq
yields another relation
\beq
\tilde{a_t} = 1 + 4\theta_t^3 \int^t_0 d^3(t,t') g(t,t')
\theta'_t dt'
\label{a}
\eeq
which completes the closed set of dynamical equations
for the system. From (\ref{gscal}) we see that the response
and the correlation functions are proportional to $\alpha^{1/4}$,
thus justifying our earlier assumption that higher-order terms
in the self-energies are smaller in $\alpha$.

We note that (\ref{se1}), (\ref{se2}) and (\ref{a})
are identical to the self-consistent equations for the $p=4$
disordered
spherical model,\cite{Crisanti93} which is particularly interesting
since there is no {\sl intrinsic} disorder in the periodic array
described by (\ref{H}) in this study. For $T > T_G$, the dynamical
equations, (\ref{se1}), (\ref{se2}) and (\ref{a}), can be collapsed
into one equation using the fluctuation-dissipation theorem and
time-translation invariance; as expected from known results
on $p >2$ spherical models,\cite{Crisanti93} there is a dynamical
instability that is {\sl unaccompanied} by a static transition.\cite{Chandra96a}
The introduction of higher order (in $\alpha$) self-energy terms
leads to dynamical equations for $p$ disordered spherical models
with several coexisting values of $p$ (i.e. p = 4,6.. where $p \ge 4$
and $p$ is even); thus this periodic array is the realization of
a ``simple spin glass'' that was previously a theorist's model
system\cite{Gross85} without a clear experimental counterpart.  We note again that the {\sl disordered} array
previously studied\cite{Vinokur87} is qualitatively similar to
the $p=2$ disordered XY model, and it is
intriguing that the absence/presence of disorder in the array leads to
the absence/presence of coinciding static and dynamical transitions.

\section{The Cooling Regimes}

The self-consistent dynamical equations, (\ref{se1}), (\ref{se2}) and
(\ref{a}),
can be simplified substantially\cite{Chandra96a} for temperatures $T > T_G$
where the response and the correlation functions depend {\sl only} on
the time-differences and can be related by the Fluctuation-Dissipation
Theorem
\beq
g(t_1 - t_2) = - \frac {\partial d(t_1 - t_2)}{\partial t_1}
\theta (t_1 - t_2).
\label{fdt}
\eeq
Here mode-coupling theory is exact because of the absence of a
length-scale
in the interaction; following previous discussions of this
approach\cite{Gotze84} we obtained\cite{Chandra96a} the closed form equation
\beq
\int_0^\infty d(t) e^{i\omega t} \theta_t dt = \frac{1 + \int_0^\infty
e^{i\omega t} d^3(t) \theta_t^4 dt}{1 - i\omega - i\omega\int_0^\infty
e^{i\omega}t d^4(t) \theta_t^4 dt}.
\label{cform}
\eeq
which results in a scaling form for the relaxation time
\beq
t_R = t_0 \tau^{-\nu}
\label{tr}
\eeq
where $\tau \equiv \frac{\theta_G - \theta}{\theta_G}$, $\nu = 1.765$ and $t_0$ is a constant that can be determined
numerically.\cite{Chandra96a}
This identification of the glass transition at $\theta_G$
assumes infinitesmally slow cooling from high temperatures;
then the system will only fail to equilibrate
when the relaxation rate diverges.

In practice the periodic array will
be subject to cooling at a finite-rate.  In this case, the
system will fall out of equilibrium
at a reduced temperature ($\theta_G^{eff} < \theta_G$) when its time-scales
associated with cooling and relaxation are equivalent,
namely when the condition
\beq
\frac{\tau}{d\tau/dt} = t_o \tau^{-\nu}
\label{eq}
\eeq
is satisfied.  (\ref{eq}) leads to the simple relation $\tau^{1+\nu} =
t_o \left(\frac{d\tau}{dt}\right) \approx t_o\left(\frac{d\theta}{dt}\right)$,
so that for a given finite cooling rate $\frac{d\theta}{dt}$
the array will be ``out of equilibrium'' for times $t > t_G^{eff}$ where
\beq
t_G^{eff} = t_0 \tau^{-\mu} = t_0 \left\{ t_0 \left(\frac{d\theta}{dt}\right)
\right\}^{-\frac{\nu}{1 + \nu}}
\label{tg}
\eeq
and undergo an effective glass transition at ${\theta_G}^{eff} \equiv
\theta(t_G^{eff})$ which is a function of its cooling rate.  In the interest
of
simplicity, we will no longer distinguish between the
system's effective and ``adiabatic'' dynamical instability; however
here we wish to emphasize that their distinction is related to
the finite nature of the cooling rate.

We would now like to characterize the array's response and correlation
functions at low temperatures ($\theta > \theta_G$) in its glassy
regime with the hope of identifying memory and history-dependent
features.
Unfortunately, in the absence of time-translational invariance
and the fluctuation-dissipation theorem, the solution of the
self-consistent dynamical equations derived in Section II.
is difficult.
As already described in the Introduction, the expected behavior
of $G_{tt'}$ and $D_{tt'}$ is qualitatively different for the
limiting cases of fast ($t \gg t_c$) and slow cooling ($t = t_c$)
regimes; here the final temperature $T_f < T_G$ is reached
at time $t_c$.
Furthermore at the onset of this study,
there were no clues about the physically interesting (and experimentally
accessible) crossover region where $t \gtrsim t_c$.
In order to clarify this situation, we now describe
our numerical studies of
the dynamical equations (\ref{se1}), (\ref{se2}) and
(\ref{a}) in the three distinct cooling regimes.
Here, in the interest of simplifying our notation,
we measure all reduced temperatures in units
of $\theta_G = \frac{3^{3/4}}{2^{1/2}}$ so that
\beq
\theta \rightarrow \theta_G \theta.
\eeq

\subsection{The Fast Cooling Regime ($ t \gg t_c$)}
 
In the limit of fast cooling all measurements are performed at times
long after the system has reached its final temperature.
Recently there has been progress in the characterization
of infinite-range (disordered) spherical models with $p$-spin
interactions in this regime;\cite{Cugliandolo93,Cugliandolo94}
these results for the $p=4$ case
are relevant for the periodic array.  Furthermore the
predicted scaling forms of the response and the correlation functions
lead to physical consequences that could be experimentally
observable; this will be discussed in Section IV. After a fast
quench, the correlation function is
predicted\cite{Cugliandolo93,Cugliandolo94} to take the form
\beq
d_{tt'} =  f \left(\frac{t' - t_0}{t - t_0}\right) d_0 \approx
\left(\frac{t' - t_0}{t - t_0}\right)^\gamma d_0
\label{Dscal}
\eeq
where
$t_0 = t_c + t_w$ is the time
when the system reaches its final temperature ($T < T_G$;
here $t_c$ and $t_w$ are the cooling and waiting times respectively.
In Figure 3 we display $d_{tt'}$ plotted as a function of
$\tilde{t} \equiv \frac{t' - t_0}{t - t_0}$ for different
cooling and measurement times where $\theta_f = 1.1$ and $0 < \tilde{t}
< 0.96$ corresponds to the ``fast quench'' regime.
Short-time dynamics
are observed in the range $\tilde{t} > 0.98$, and $\tilde{t} < 0$
corresponds to the ``time range'' when the temperature was changing
at a finite rate; this range vanishes in the limit of
very fast cooling.  In Figure 3, we see numerical
support for the scaling of the correlation function
implied by expression (\ref{Dscal}); we see there
that with increasing $t$, the correlation
functions corresponding to finite-cooling rate converge
to that for infinitely fast cooling.
A log-log plot of the $d_{tt'}$ for the ``fastest''
response (with the least ``tail'' for $\tilde{t} < 0$) is linear,
and yields
$\gamma =0.43$ and $d_0=0.8$ for $\theta_f = 1.1$,
consistent with the numerical trends previously
reported for the $p$-spin spherical
models.\cite{Cugliandolo93,Cugliandolo94}
In the case of infinitely fast cooling $t_0 = t_w + t_c$,
where $t_w$ is the waiting time at the initial (high) temperature.
Furthermore we found that the correlation function
for finite-rate cooling is well-described by the scaling
form (\ref{Dscal}) even for relatively slow cooling rates
if $t_0$ is adjusted so that it corresponds to the
time at which the system falls out of equilibrium; now
$t_0^{eff} = t_w + t_c - \delta t$.

\centerline{\epsfxsize=8cm \epsfbox{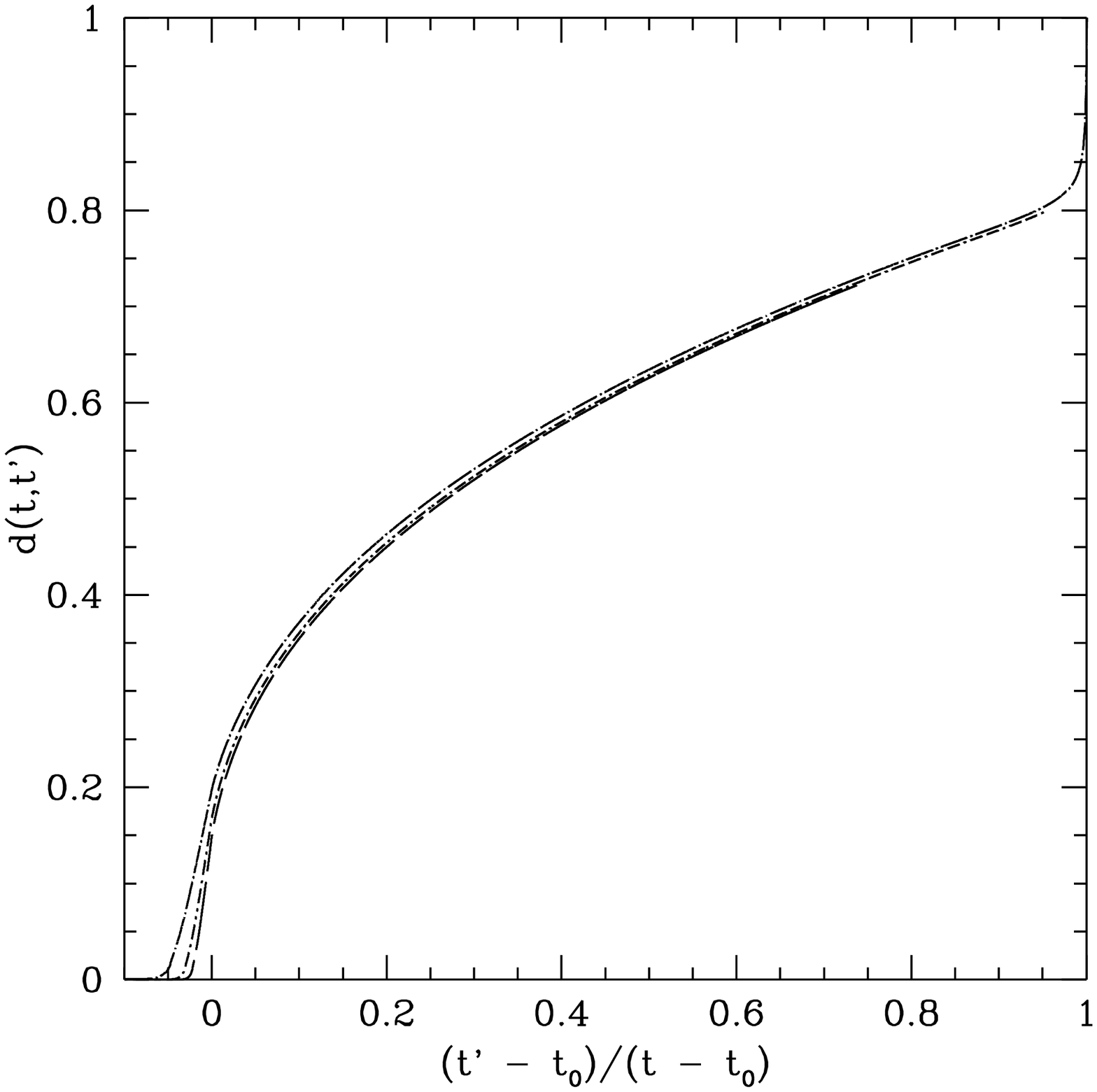}} 
{\footnotesize {\bf Fig~3. Numerical evidence for scaling of the
correlation function $D_{tt'}$ after a fast quench to reduced
temperature $\theta_f = 1.1$.} The solid
line corresponds to the case of an almost infinitely fast quench.
The other curves correspond to a finite-cooling rate of
$\frac{d\theta}{dt} = 0.05$  to the same reduced temperature
with different measurement times t=200,300 and 400 (top to bottom);
we observe that these finite-rate correlations converge to
the fast quench case with increasing $t$.
}\vspace{0.5in} 

A generalized fluctuation-dissipation theorem (gFDT)
\beq
x = \frac{g_{tt'}}{d'_{tt'}}
\label{qfdt}
\eeq
has also been identified\cite{Cugliandolo93,Cugliandolo94}
in this cooling limit for the range
$0 < \tilde{t} < 1$.  In Figure 4 the numerical results
strongly support this proposal (note that $\tilde{t} >
0.8$ and
$\tilde{t} \sim 0$ correspond to short-time dynamics and to
time-interval
where the temperature is varying); furthermore
the value for $x = 0.55$ is in close agreement with that expected
from the analytical treatment\cite{Cugliandolo93,Cugliandolo94}
($x = {(p-2)(1-q)}$ with
$p=4$ and $q = \frac{2}{3}$). We also note that the scaling
form (\ref{Dscal}) appears to be robust to decreasing
final temperature, though the
specific values of $d_0$ and $\gamma$ change
(for $\theta_f = 1.3$, $d_0 = 0.95$ and $\gamma = 0.2$ in contrast
to the values $d_0 = 0.8$ and $\gamma = 0.43$ for $\theta_f = 1.1$).

\centerline{\epsfxsize=8cm \epsfbox{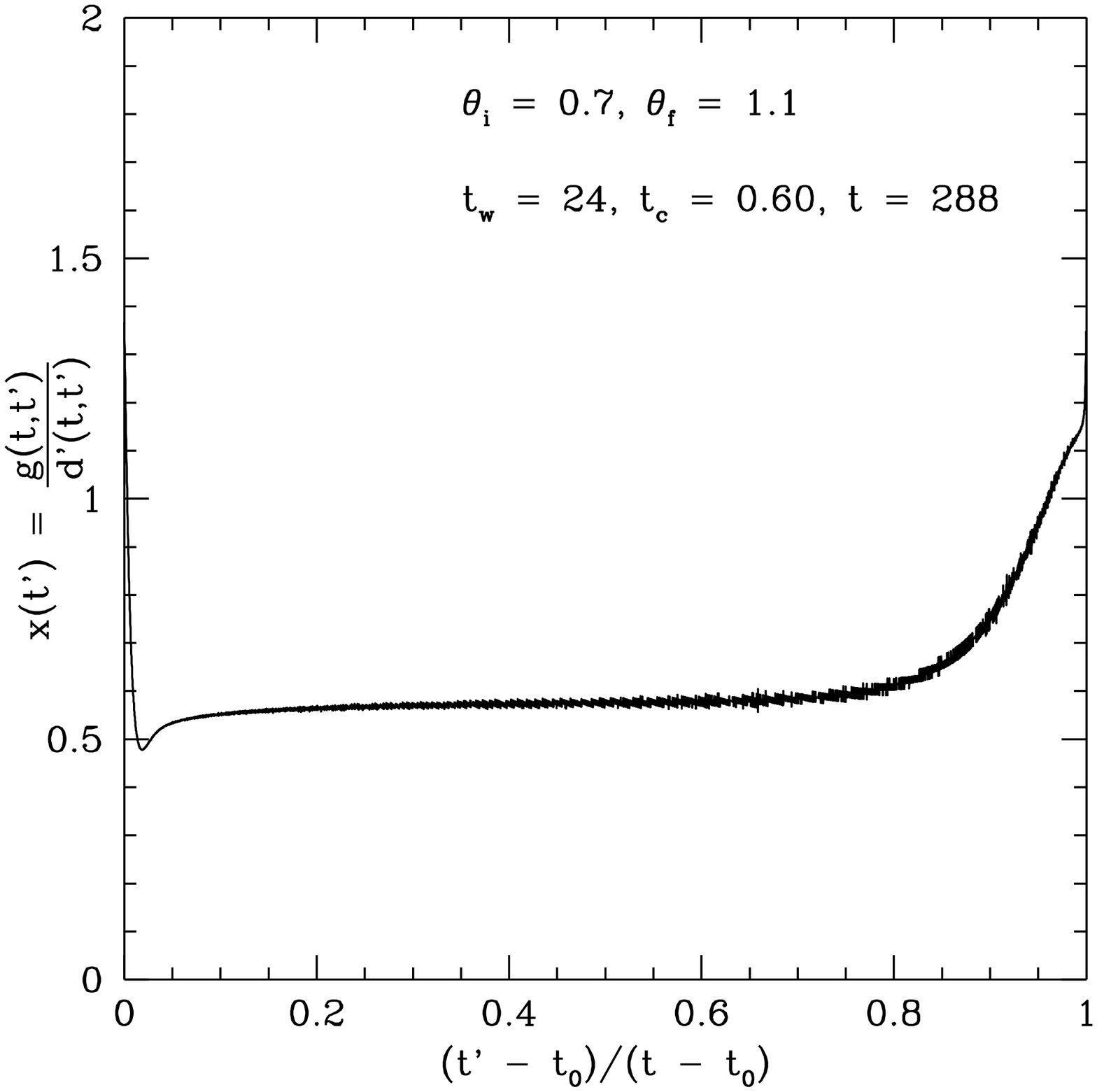}} 
{\footnotesize {\bf Fig 4. Numerical support for the generalized FDT relation
after a fast quench.} Here we plot $x(t') = {g(t,t')\over d'(t,t')}$
as a function of $\tilde{t} = (t' - t_0)/(t - t_0)$ which is
constant but less than unity for $\tilde{t} < 0.8$ where
$t_0$ corresponds to the time when there is no longer temperature
variation; $\tilde{t} > 0.8$
corresponds to short-time dynamics.
}\vspace{0.5in} 

\subsection{The Slow Cooling Regime ($t = t_c$)}

The other limiting case is that of slow cooling where the
measurement is taken at the final cooling time ($t = t_c$).
We expect $t^*$, a time-scale that is a function of the
cooling rate $\Gamma = \left(\frac{d\tau}{dt}\right)$, to play a key
role in the time-dependence of the response and correlation
functions; more specifically the latter will be a function
of $\frac{t'}{t^*}$ and $\frac{t}{t*}$ as displayed in Table
I. where $t^* = f\left(\frac{1}{\Gamma}\right)$. The form
of the functional dependence of $t^*$ on the cooling rate
must be determined numerically.  In Figure 5, we plot
$\frac{d_{tt'}}{\Gamma}$ and $g_{tt'}$ vs. $\Gamma t'$
where $\Gamma = \left(\frac{d\tau}{dt}\right)$ for $\theta_f = 1.1$.
We remark that in the absence of anomalous scaling ($\eta = 0$),
$d_{tt'}$ and $g_{tt'}$ are simply functions of the reduced
temperatures $\theta_t$ and $\theta_t'$ so that in this
regime they can be plotted in terms of $\tau = \theta_t' - 1$.
We observe
that the characteristic width decreases with
increasing cooling times suggesting the need for another
functional form for $\Gamma$.  In Figure 6, we present another
scaling plot for the same data as in Figure 5 ($\theta_f = 1.1$); this
time we plot $\frac{d_{tt'}}{\tilde\Gamma}$ and $g_{tt'}$
vs. $\tilde\Gamma t'$
where $\tilde{\Gamma} = \Gamma^{1 - \eta}$ with $\eta = 0.19$; now
the curves fall on top of one another nicely.
In this case $d_{tt'}$ and $g_{tt'}$ are {\sl not} just functions
of the reduced temperatures.
Scaling
plots are also presented for data associated with $\theta_f = 1.2$
and $\theta_f = 1.3$ in Figures 7 and 8; we infer from these
results that
\beq
t^* = \left(\frac{1}{\Gamma}\right)^{1 - \eta(T)}
\label{t*}
\eeq
where $\eta$ decreases with decreasing temperature.
More specifically we find that $\eta = 0.19, 0.12$ and $0.0$
for $\theta_f = 1.1, 1.2$ and $1.3$, and the
associated scaling plots are displayed in Figures 6 - 8.
In Figures 6 and 7 $\eta \neq 0$ and there is a new
time-scale,
$t^*$, associated with the decay of the correlation
functions.  The scaling plots in Figure 8 indicate
that $\eta = 0$ at lower temperatures,
suggesting that the array ``rejuvenates''
and improves its memory there.  We note that
$\eta = 0$ in the slow cooling regime of the Sherrington-Kirkpatrick
model;\cite{Ioffe88} it is somewhat amusing that
such ageing effects,\cite{Vincent96,Fischer91} well-known in spin
glasses,
should
appear in a model for a periodic array. The observable consequences
of this ``forgetfulness'' and subsequent ``rejuvenation''
will be discussed in Section III.
Finally we note that at temperatures such that
$\eta = 0$ $\frac{\partial d(t,t')}{\partial t'}$
is constant for a wide temperature range ressembling
the slow cooling solution of the Sherrington-Kirkpatrick
model\cite{Ioffe88}. We also recall that there were no qualitative
changes in the fast-cooling regime as a function of increasing
$\theta_f$ (decreasing temperature).

\centerline{\epsfxsize=8cm \epsfbox{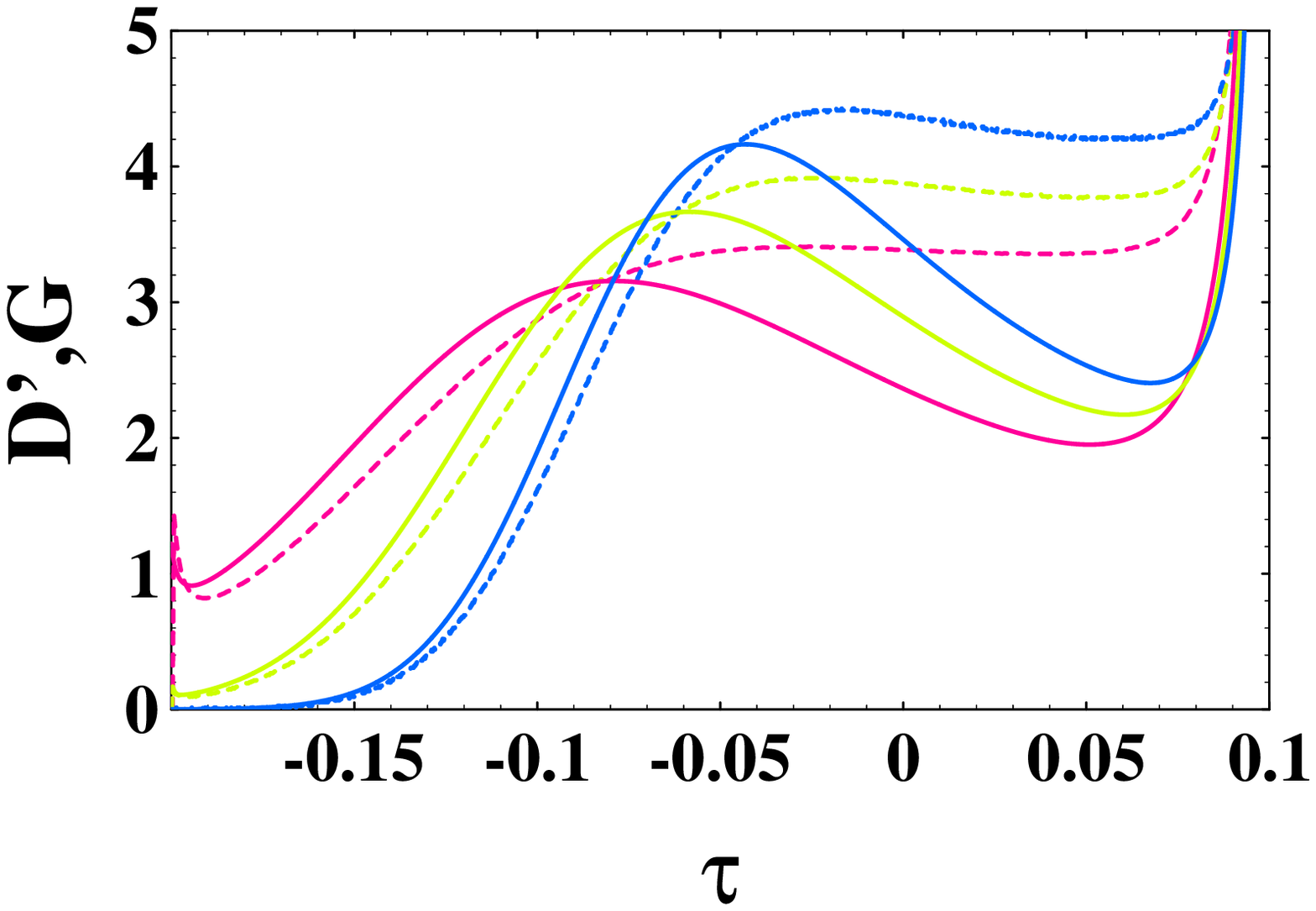}} 
\vspace{-1.0in}

{\footnotesize {\bf Fig 5. Absence of the adiabatic limit in the slow
cooling regime.}  Here we plot $\frac{1}{\Gamma}\frac{\partial d(t,t')}
{\partial t'}$ and $\frac{1}{\Gamma}g(t,t')$
vs. $\tau=\Gamma (t'-t_G)=\theta(t')-1$ for different cooling rates ($\Gamma =
0.3/40, 0.3/80$ and $0.3/160$); the characteristic width of the curves
decreases with increasing cooling times. All dashed curves are
$d'(t,t')$ and the solid ones are $g(t,t')$, and the cooling time
increases from bottom to top.
}\vspace{0.5in} 

\centerline{\epsfxsize=8cm \epsfbox{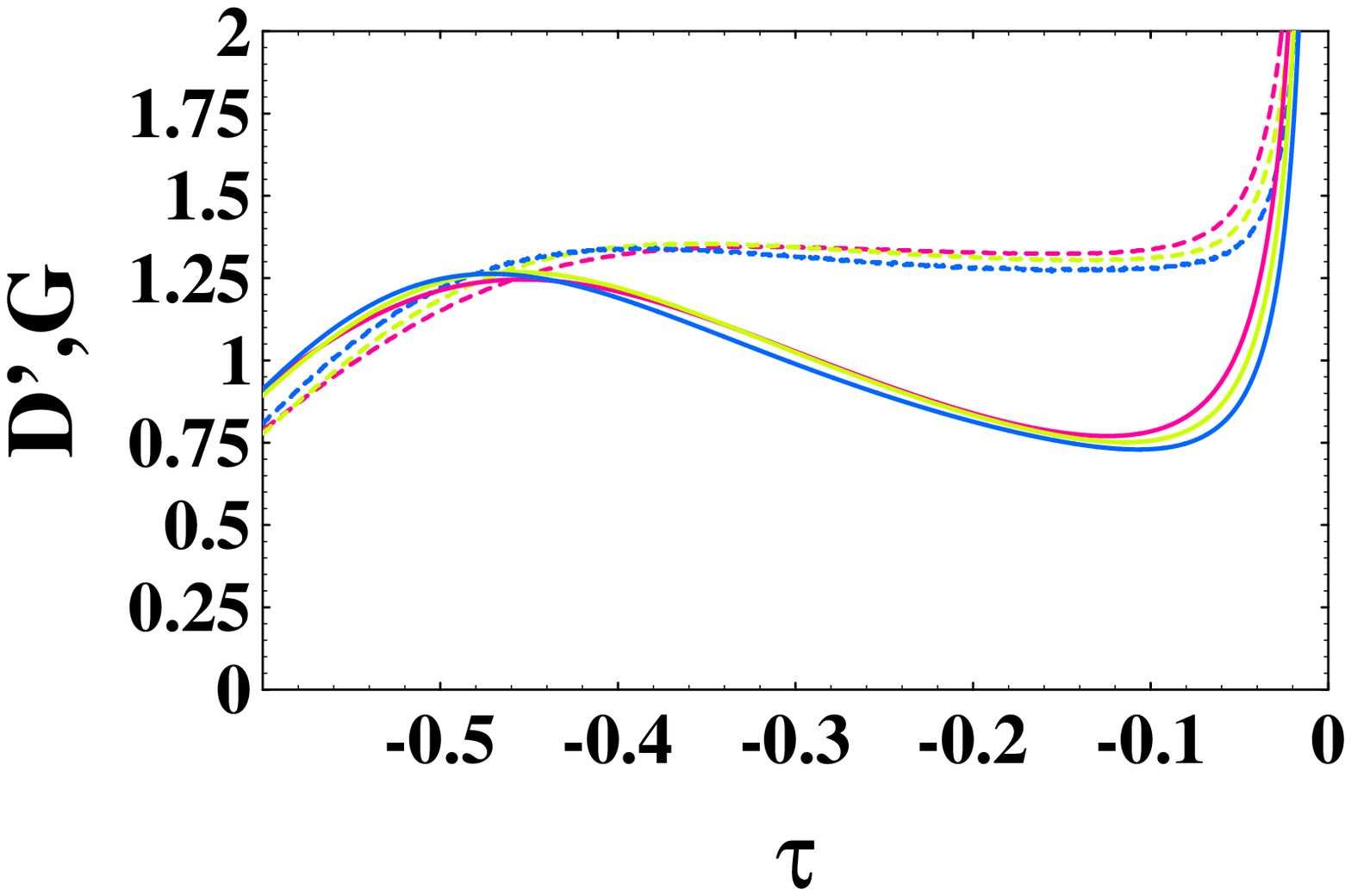}} 
\vspace{-1.0in}

{\footnotesize {\bf Fig 6. A scaling plot of $\frac{1}{\tilde\Gamma}
\frac{\partial d(t,t')}{\partial t'}$ vs $\tau(t')=\tilde{\Gamma}(t'-t)$} for
three different cooling rates ($\Gamma = 0.3/40, 0.3/80$ and $0.3/160$)
where
$\tilde{\Gamma} = \Gamma^{1 - \eta}$ with $\eta = 0.19$; here
$\theta_f = 1.1$ All dashed curves are
$d'(t,t')$ and the solid ones are $g(t,t')$.
}\vspace{0.5in} 

\centerline{\epsfxsize=8cm \epsfbox{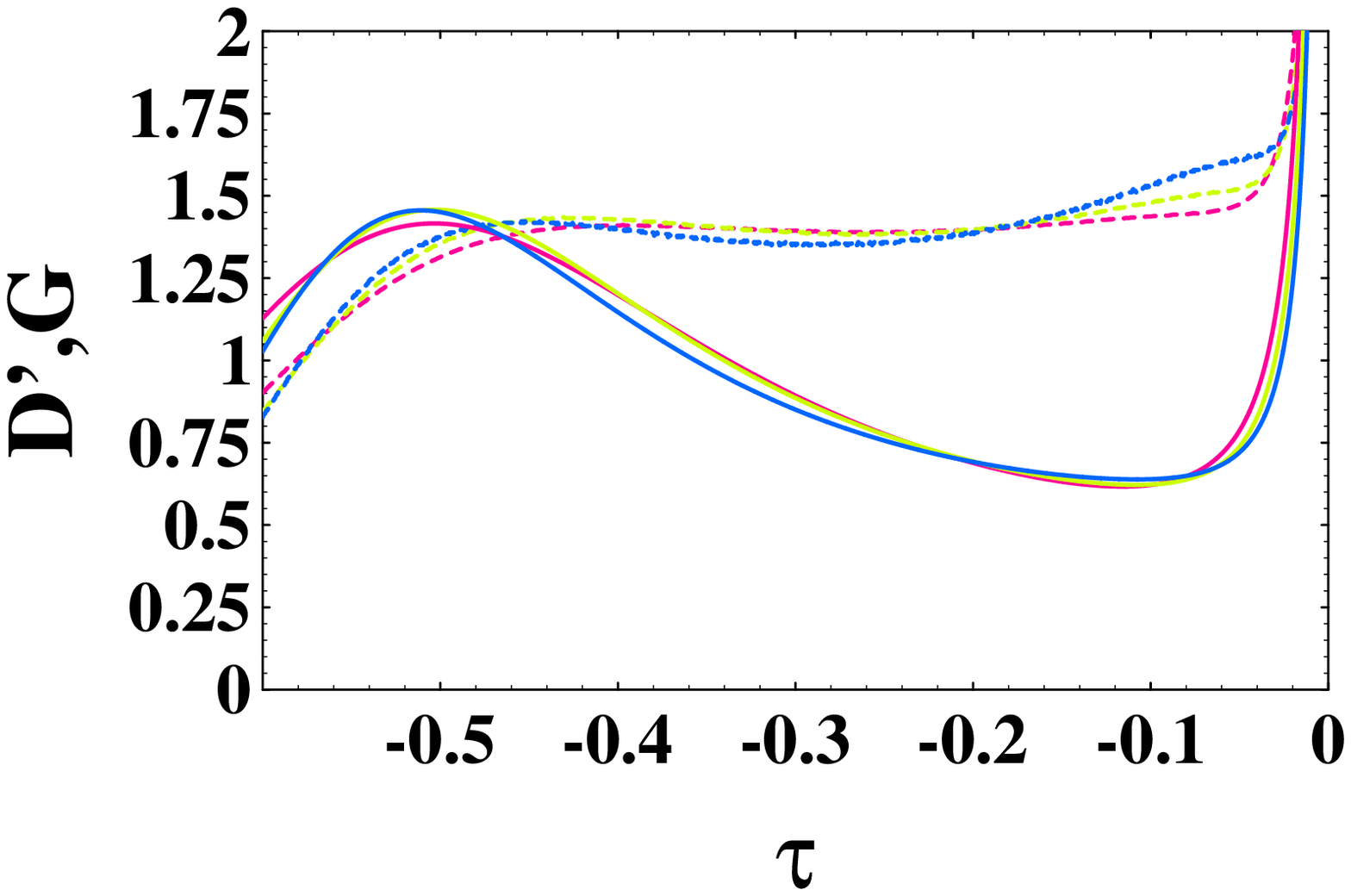}} 
\vspace{-1.0in}

{\footnotesize {\bf Fig 7. A scaling plot of $\frac{1}{\tilde\Gamma}
\frac{\partial d(t,t')}{\partial t'}$ vs $\tau(t')=\tilde{\Gamma}(t'-t)$} for
three different cooling rates  ($\Gamma = 0.3/40, 0.3/80$ and $0.3/160$)
where $\tilde{\Gamma} = \Gamma^{1 - \eta}$ with $\eta = 0.12$;
here $\theta_f =1.2$. All dashed curves are
$d'(t,t')$ and the solid ones are $g(t,t')$.
}\vspace{0.5in} 

\centerline{\epsfxsize=8cm \epsfbox{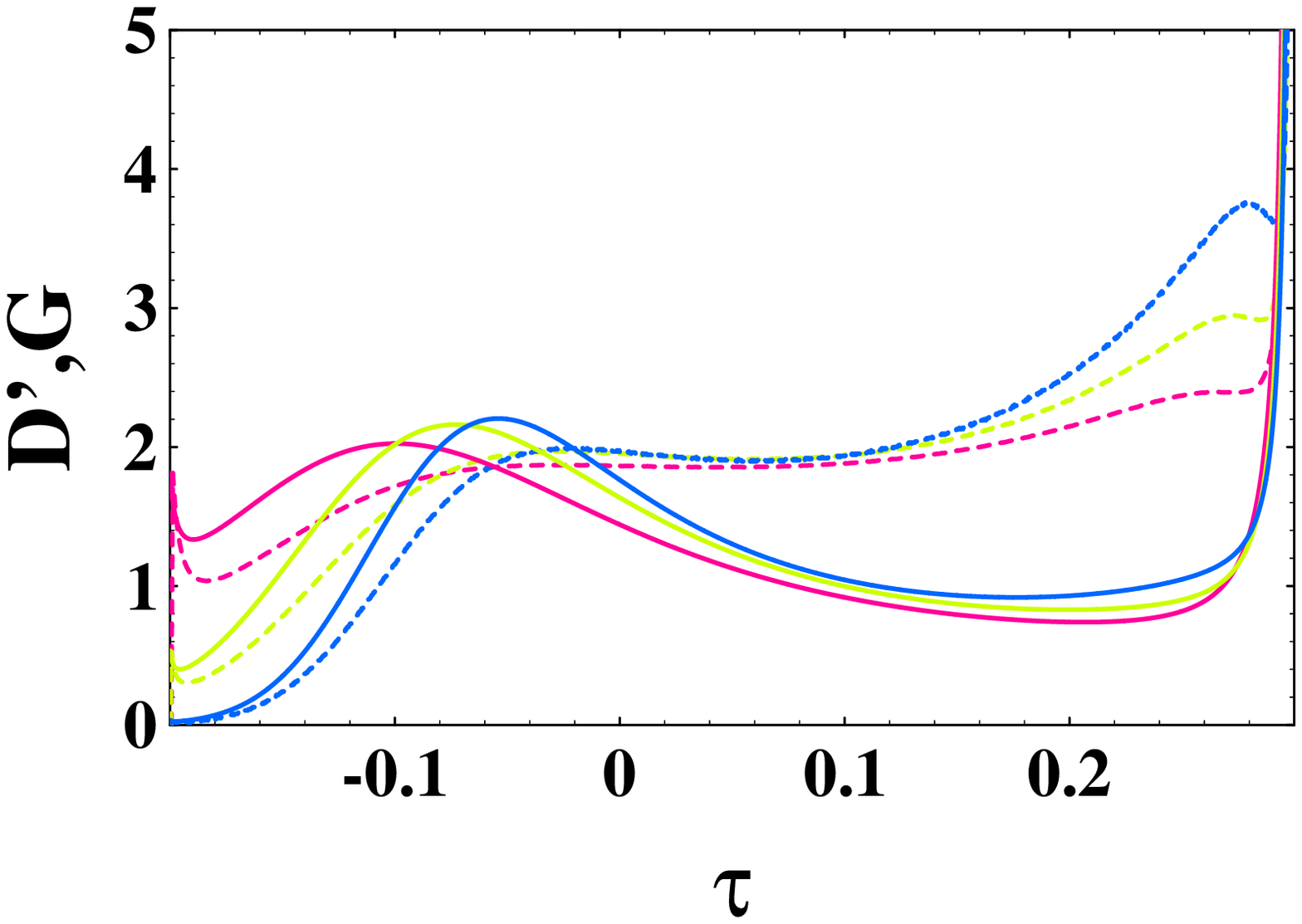}} 
\vspace{-1.0in}

{\footnotesize {\bf Fig 8. A scaling plot of
$\frac{1}{\tilde\Gamma}\frac{\partial d(t,t')}{\partial t'}$ vs
$\tau(t')=\Gamma (t'-t_G)=\theta(t')-1$}
for three different cooling rates ($\Gamma = 0.3/40, 0.3/80$ and $0.3/160$)
indicating the appearance of the adiabatic limit at $\theta_f = 1.3$.
All dashed curves are
$d'(t,t')$ and the solid ones are $g(t,t')$, and the cooling time
increases from bottom to top.

}\vspace{0.5in} 

\subsection{The Crossover Cooling Regime ($t > t_c$)}

As indicated in Table I, we had no initial expectations for
the form of $d_{tt'}$ and $g_{tt'}$ in the intermediate cooling
regime $t \gtrsim t_c$.  In Figure 9 we display a number of correlation
functions that display crossover behavior between the two
limits of the fast and slow cooling cases.
Here we see that at $t \gtrsim t_c$ there exist at least two
distinct regimes:  $t' <
t_0$
and $t' > t_0$.  In the first regime $\frac{\partial d}{\partial t'}$ is approximately constant,
whereas in the second case the correlation function varies more like
a power-law ressembling the fast cooling case. In order
to emphasize this point, we have included a scaling curve
$D(t') = d_0 \left(\frac{\tilde{t} - x}{1 - x}\right)^\gamma$ where
$\tilde{t} = \frac{t' - t_0}{t - t_0}$; here $x=0.13$ is parameter
that effectively rescales $t_0$,
and we have taken the parameters $d_0 = 0.8$ and
$\gamma = 0.43$ from the fast-cooling case (see Figure 3) where
$\theta_f = 1.1$. We note that the scaling curve and the correlation
function ($t=432$) agree very well for $\tilde{t} > 0.3$;
we also see an intermediate regime $0 < \tilde{t} < 0.3$ where there is
still curvature in $D(t')$ but no agreement with the scaling
form (\ref{Dscal}).
The generalized FDT ratio $x$, displayed in Figure  10,
is as previously observed for fast cooling (cf. Figure 4) for
$\tilde{t} = \frac{t' - t_0}{t - t_0} > 0$, but is no longer
constant for $\tilde{t} < 0$ (slow cooling).  By contrast in Figure 11
we observe that $\frac{\partial d(t,t')}{\partial t'}$ is almost constant
as in the slow-cooling regime discussed above (see Figs 5-7).
This overall behavior suggests that for ``crossover measurement times''
($t \gtrsim  t_c$), the system's behavior  is a combination of that
observed for both fast and slow cooling. As discussed in Section IIB,
the behavior of the slow-cooling correlation function changes
qualitatively when the final temperature is sufficiently low
($\theta_f \ge 1.3$).  This change also affects the correlation
function in the crossover regime.  As we observe in Figure 12 the
$\frac{\partial d(t,t')}{\Gamma \partial dt'}$ acquires a peak
and, more importantly, it ceases to depend on the cooling rate
$\Gamma$
indicating that the system preserves its memory of the cooling process
even as $t \rightarrow \infty$ (with the ratio $\frac{t}{t_c}$
maintained
constant). We note that the plateau displayed in Figure 12 corresponds
to that of $d'$ shown in Figures 7 and 8 (in the slow cooling regime).
The range $\tilde t = \frac{t' - t_0}{t - t_0} > 0$ corresponds
to the fast cooling regime where $\frac{\partial d}{d t'} \sim
\frac{1}{t'}^{\gamma - 1}$;
the observed peak is needed to match the fast- and slow- cooling
regimes.

\centerline{\epsfxsize=8cm \epsfbox{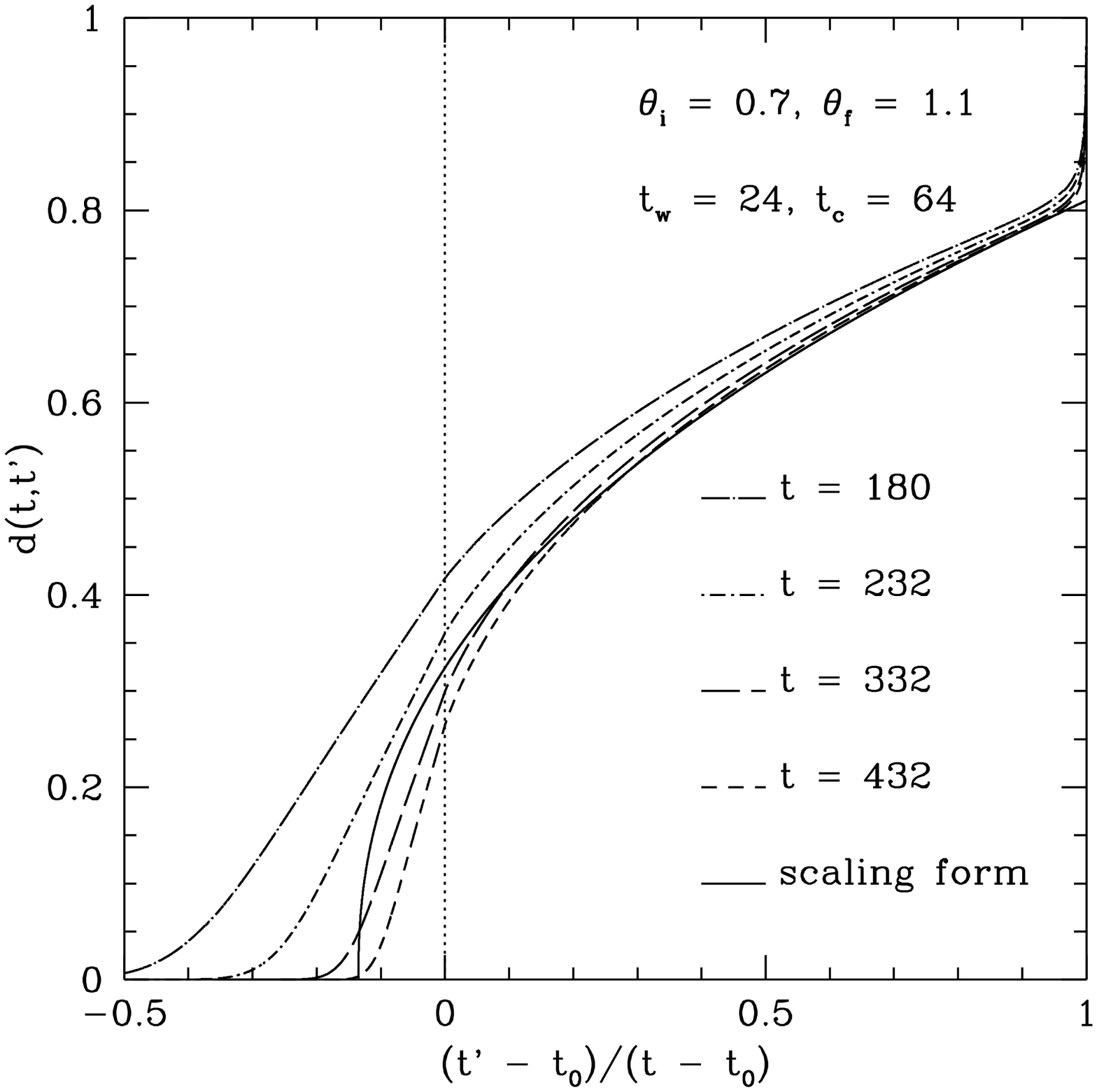}} 
{\footnotesize {\bf Fig 9. The correlation function $d(t,t')$ in the crossover
between fast and slow cooling regime for cooling rate $\Gamma = 0.0125$ and
$t=180,232,332,432$  and $\theta_f = 1.1$ (top to bottom).}
Note the presence of two different regimes evident for $t \sim t_c$ : (1)
$\frac{\partial
d(t,t')}{\partial t'} \approx$ constant for $\tilde{t} < 0$ and (2)
power-law variations at $\tilde{t} > 0$ ressembling the fast cooling
regime where $\tilde{t} = \frac{t' - t_0}{t - t_0}$.
Note that at $t$ becomes larger than $t_c$ the linear slow-cooling
regime rapidly vanishes and that the correlation function converges
to the scaling form (\ref{Dscal}) (dash-dot curve)
$D = d_0 \left(\frac{\tilde{t} - x}{1 - x}\right)^\gamma$
with $d_0 = 0.8$, $\gamma = 0.43$ and $x = -0.13$.
}\vspace{0.5in} 

\centerline{\epsfxsize=8cm \epsfbox{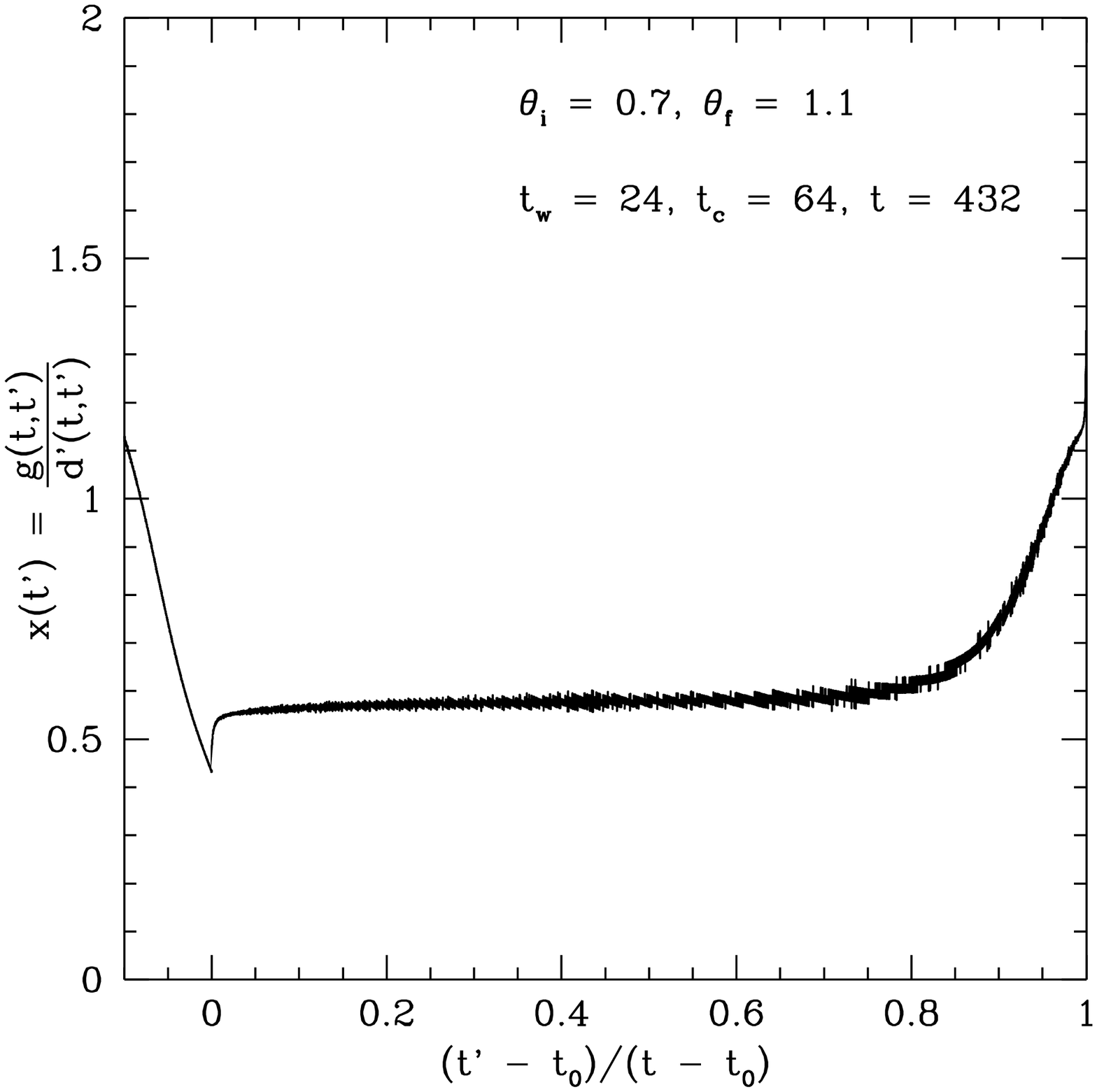}} 
{\footnotesize {\bf Fig 10. The generalized FDT ratio $x$ in the crossover
regime between slow
and fast cooling for $\Gamma = 0.125$, $t = 432$ and $\theta_f = 1.1$.}
Note that in this plot $t$ was sufficiently larger than $t_c$
and the modified FDT relation is maintained for most
of the range ($(t' - t_0)/(t - t_0) > 0$); furthermore  $x$ is
as previously observed for fast cooling, but for $(t' - t_0)/(t - t_0) <
0$ the ratio is no longer constant.
}\vspace{0.5in} 

\centerline{\epsfxsize=8cm \epsfbox{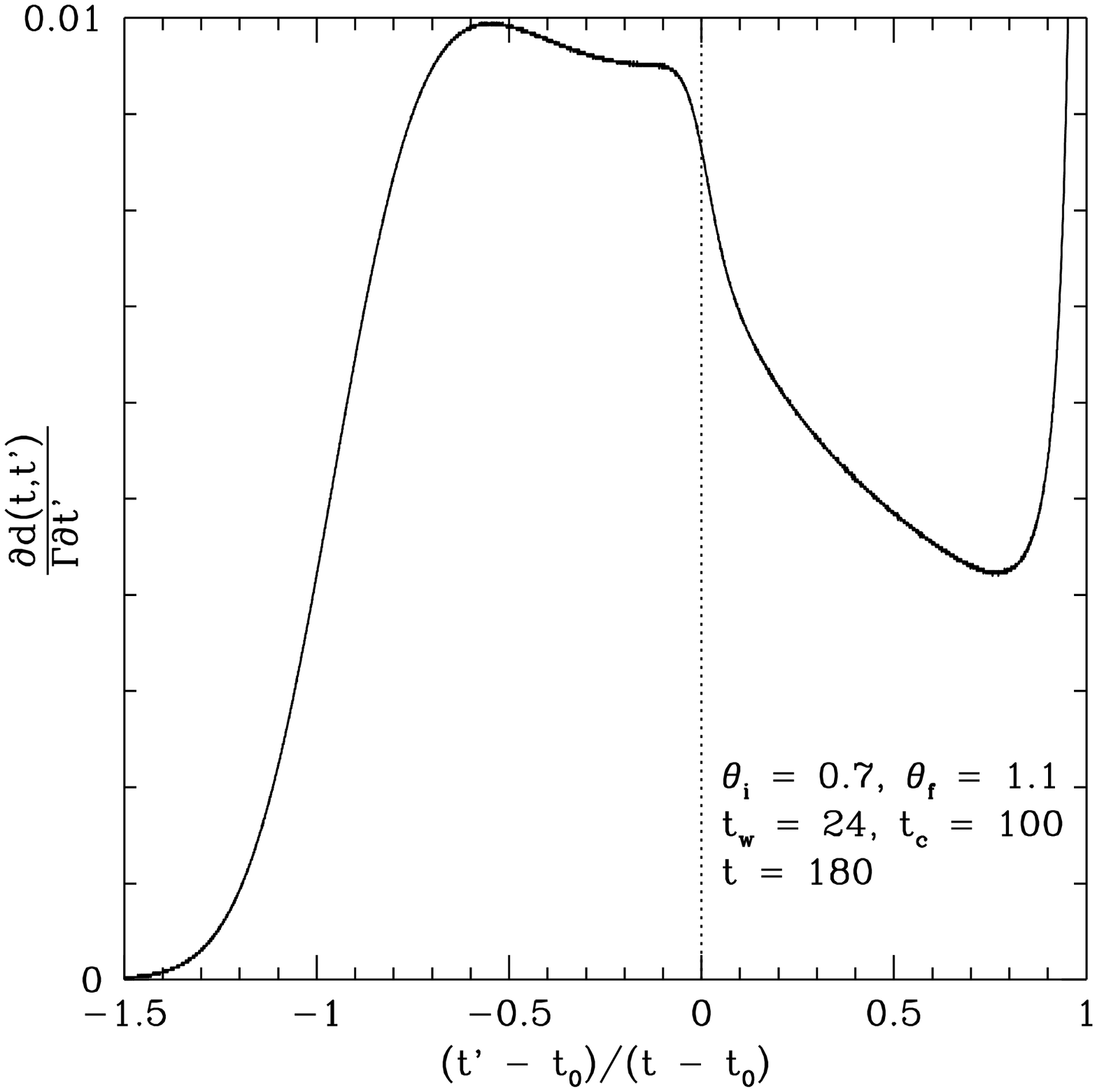}} 
{\footnotesize {\bf Fig 11. The derivative of the correlation function,
$\frac{\partial
d(t,t')}{\Gamma\partial t'}$ for $\Gamma = 0.004$, $t = 180$ and
$\theta_f = 1.1$ in
the crossover between slow and fast cooling.}
In this figure $t \sim t_c$ and we see
the appearance of the behavior characteristic
of slow cooling ($\frac{\partial
d(t,t')}{\partial t'} \approx$ constant).
}\vspace{0.5in} 

\centerline{\epsfxsize=8cm \epsfbox{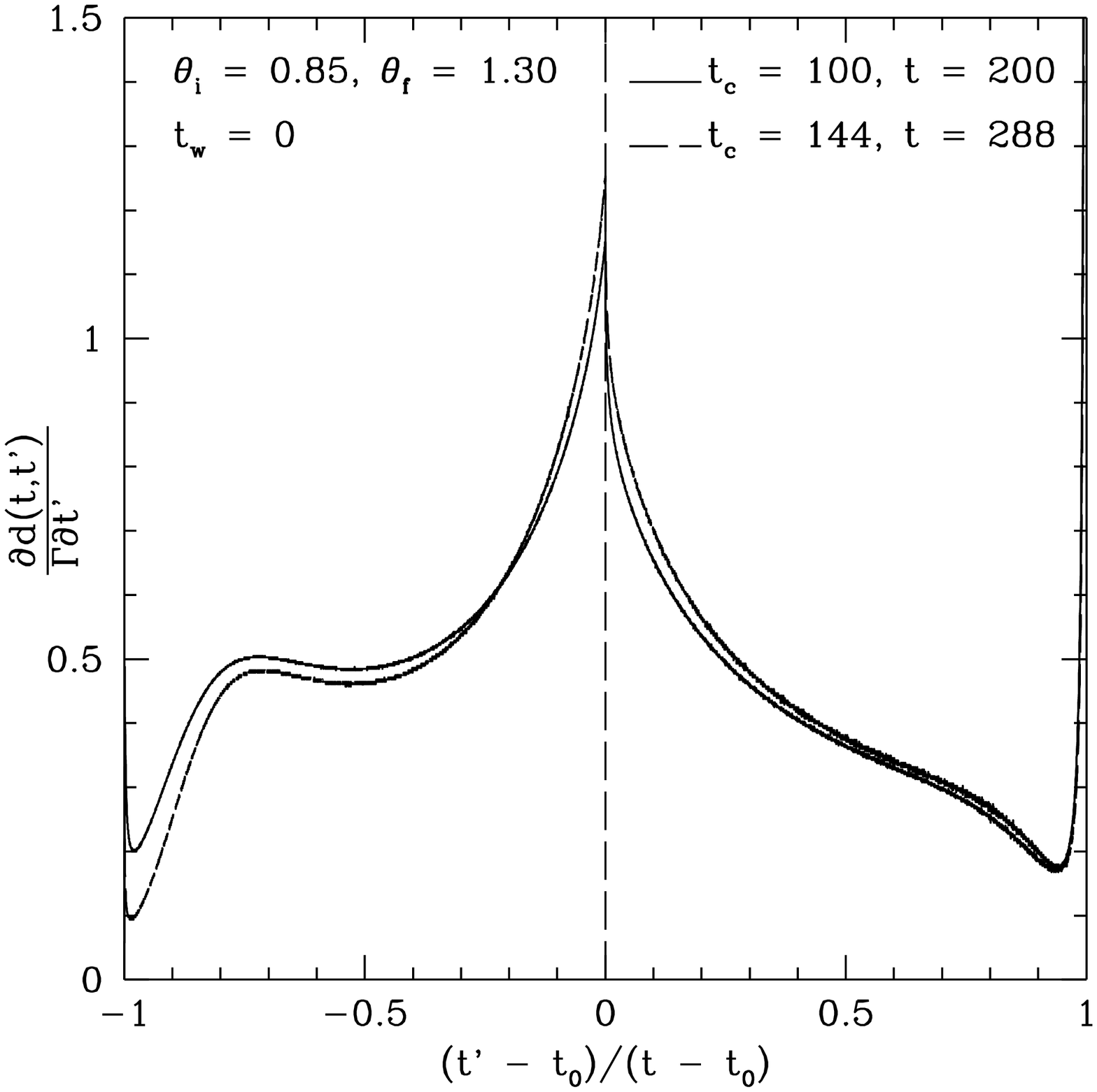}} 
{\footnotesize {\bf Fig 12. The derivative of the correlation function,
$\frac{\partial d(t,t')}{\Gamma\partial t'}$ for two
cooling rates, $\Gamma = 0.45/100, .45/144$
with the {\sl same} ratio $\frac{t}{t_0} = 2$ and $\theta_f = 1.3$.}
We see that $\frac{\partial d(t,t')}{\Gamma\partial t'}$ is constant
even in the slow cooling regime indicating ``perfect'' memory.
}\vspace{0.5in} 

\subsection{Non-Monotonic Cooling and Structure of the Phase
Space}

It is known that the structure of the metastable states in
the (disordered) Sherrington-Kirkpatrick (SK) model is rather
complex; the states are arranged on an ultrametric tree
with many branch points.
Here we provide evidence that the situation is rather different
for the periodic array discussed here. In the glassy phase of
the SK model,  each metastable state subdivides and
``daughter'' states appear as the temperature
is lowered.  In this case application of a heat pulse removes
some of this subdivision and results in a partial loss of memory.
For the periodic array such behavior is {\sl not} observed but,
by contrast, a small heat pulse does not lead to any aftereffect.
Of course a large heat pulse brings the system to above its
effective glass transition temperature and thus erases {\sl all}
of its memory.  In Figure 13 we display the correlation functions
for varying amplitudes of applied heat pulses.  Here the system
was cooled fast to low temperature ($\theta = 1.4$), equilibrated
and then subjected to a heat pulse.  From the results shown in
Figure 13 we conclude that a small heat pulses does not change the
state of the array at $\theta \ll \theta_G$,
while a large one brings it above the
effective glass transition temperature and eliminates
all memory of its initial state. If the metastable states
did subdivide with decreasing temperature, the correlation
and response functions obtained with and without the 
application of
a moderate heat pulse would
be different; in the former case memory of some of
the subdivisions would be erased.  Clearly
this is not the case for the periodic array,
and 
response functions for this cooling-heating process
(not shown) are also consistent with 
this conclusion. 
Furthermore, we checked that the results
obtained when the system was cooled slowly to $\theta=1.4$,
equilibrated
and then subjected to the heat pulse were qualitatively similar.
>From these results we conclude that there is {\sl no} subdivision
of metastable states below the glass transition temperature.

\centerline{\epsfxsize=8cm \epsfbox{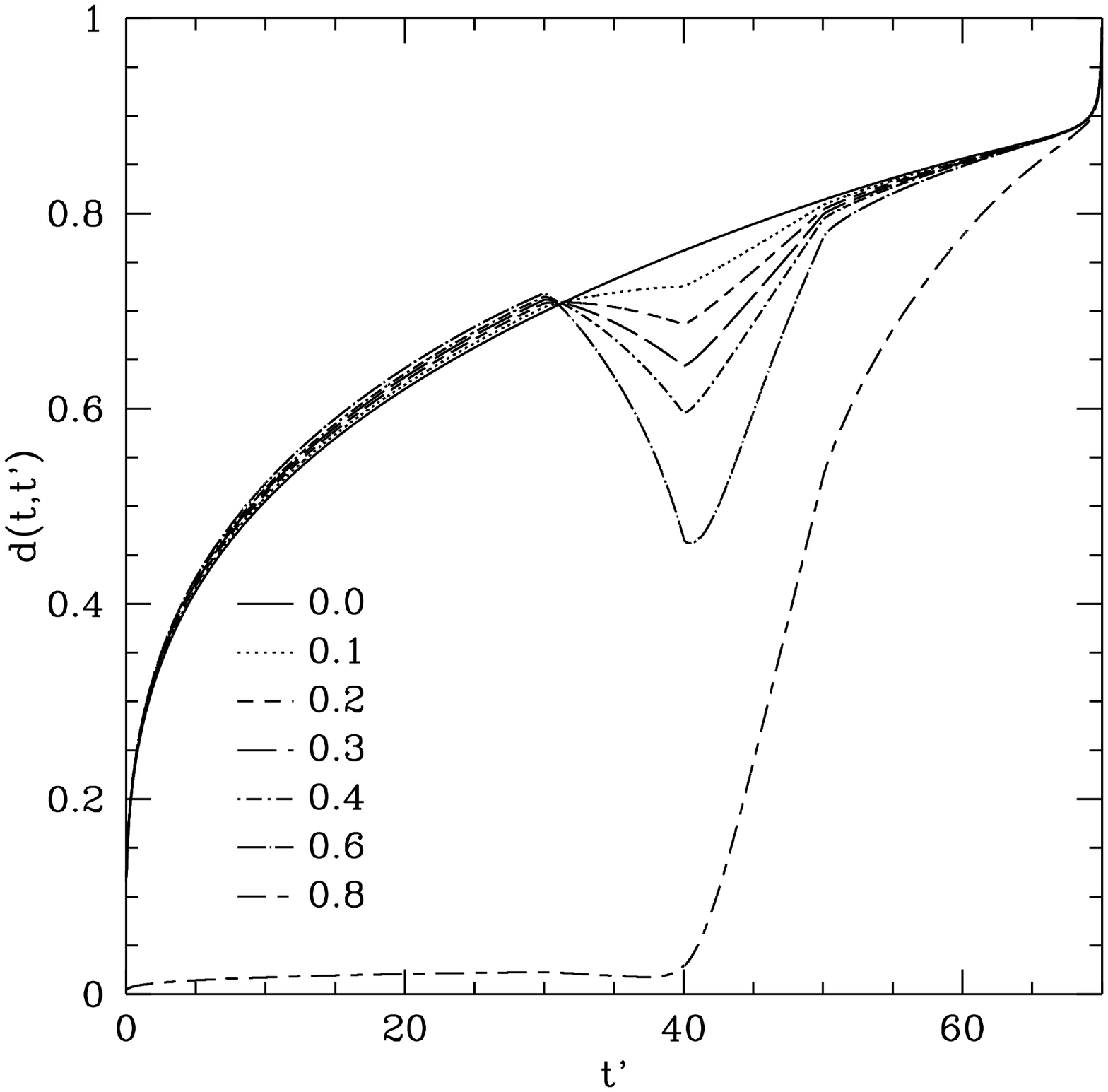}} 
{\footnotesize {\bf Fig 13. The correlation function, $d(t,t')$, for the
cooling-heating regime}.
Here the temperature was reduced infinitely fast to $\theta=1.4$
at $t' = 0$;
the system was equilibrated for $0 < t' < 30$, then heated linearly
for $30 < t' < 40$, then cooled linearly back to $\theta = 1.4$ for
$40 < t' < 50$ and then measured at $t = 70$. Different curves
correspond to different amplitudes of the heat pulse as shown.
We see that for heat pulses of amplitude less than $0.6$ the
system recovers its original state; however for larger heat pulses
all memory is lost.}
\vspace{0.5in} 

We also checked that these results were not due to the leading-order
approximations made in the derivation of the dynamical equations,
and thus performed direct simulations of the array to
confirm that this conclusion remains valid at low temperatures
and at large $\alpha$.  First we simulated finite-size arrays
with $\alpha = 1.0$ at very low temperatures and confirmed that
the number of metastable states
grows exponentially with system-size. We also studied the distribution
of overlaps between these states, and our results are displayed
in Figure 14 for $\alpha = 1.0$. 
Here we see that the overlap distribution evolves to a sharp peak
with a maximum that shifts to lower $q$ with increasing number of 
wires $N$.  In particular we do not observe any states with
large overlaps for $N \geq 15$; furthermore $P(q)$ suggests that
for large values of $N$ the states are equidistant in phase space.
We also found qualitatively similar results for smaller values of
$\alpha$.

\centerline{\epsfxsize=8cm \epsfbox{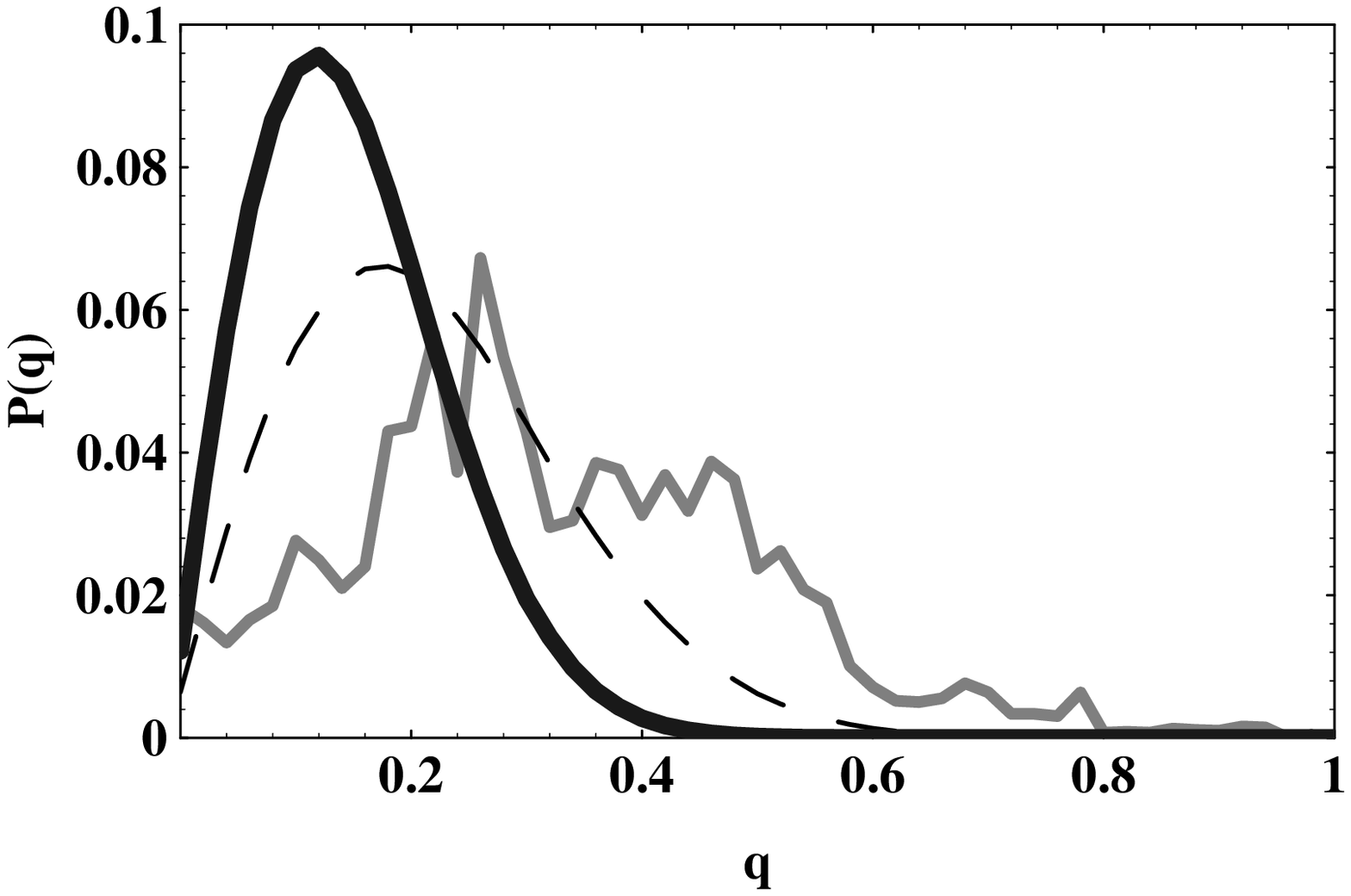}} 
{\footnotesize {\bf Fig 14. Distribution of overlaps for finite-size
arrays with $\alpha = 1.0$}. Here the gray, dashed and black curves
correspond to $N = 7, 15$ and $31$ respectively. }
\vspace{0.5in} 

The equidistance of metastable states does not imply that the
structure of the phase space is simple. In particular
the basins of attraction associated with different
metastable states appear to be complicated.  For example,
we have studied the probability, $Q(q)$, that a randomly chosen state that has
overlap
$q$ with a metastable one evolves to this particular metastable state.
The results are displayed in Figure 15.  We see that the probability
to evolve to {\sl different} states remains finite even if the
starting
overlap is close to unity. Our results suggest that $1 - Q(q) \propto
(1 - q)$ for $N \rightarrow \infty$. 
If true, this has important consequences for the distribution
of barriers. First we note that  
the energy required to flip
$m$ spins cannot be larger than $mT_0$.
The probability of evolving to a different metastable
state after the flipping of $m$ spins is roughly $\frac{m}{N}$
so the probability to find {\sl some} $m$ spins which lead
to this result is $O(1)$.  This implies that even in the
infinite-range periodic 
system there is a broad distribution of barriers which
extend down to $E_B \sim T_0$.
We note that this result implies that the distribution
of barriers and overlaps are {\sl not} related.

\centerline{\epsfxsize=8cm \epsfbox{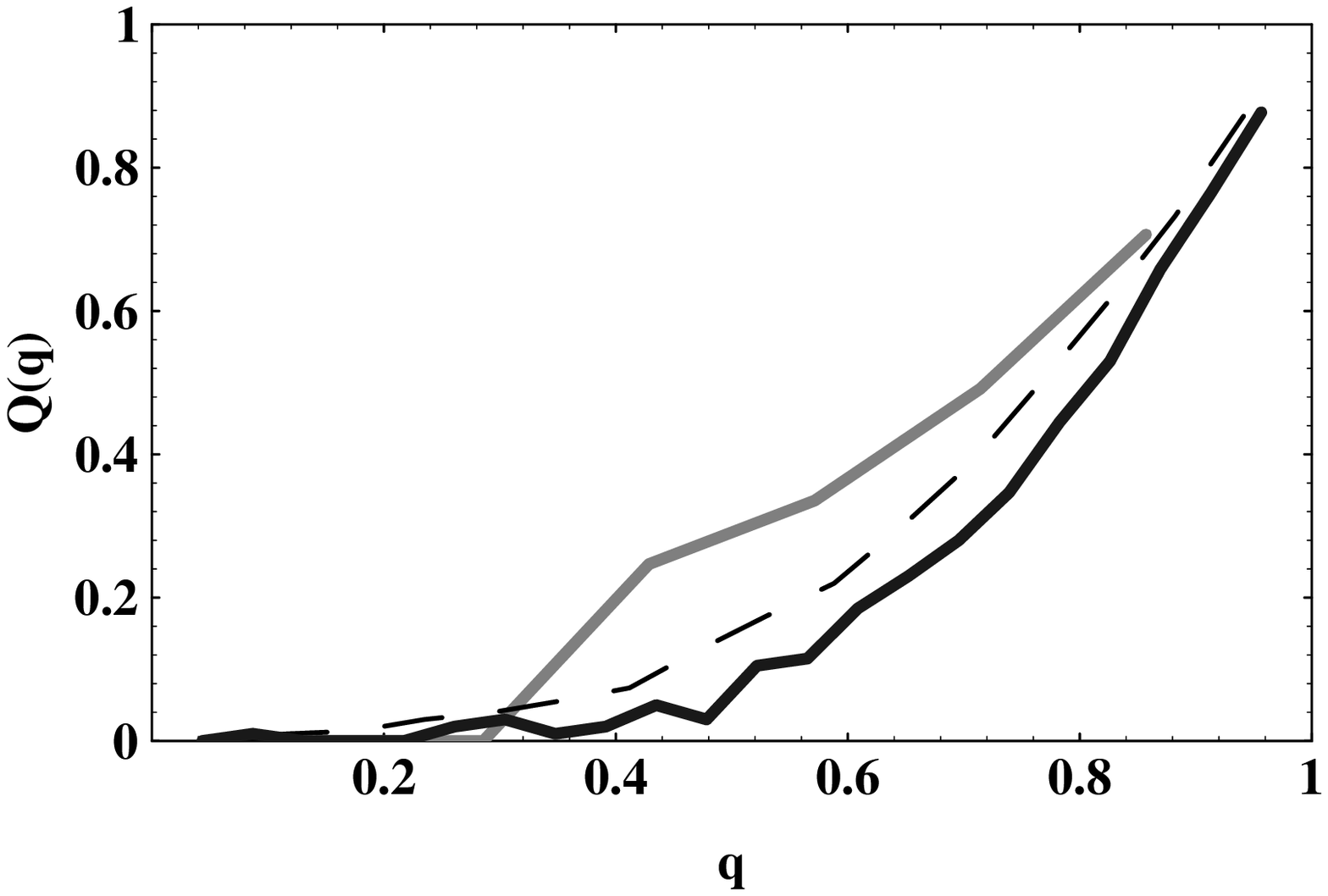}} 
{\footnotesize {\bf Fig 15. The probability, $Q(q)$, to evolve
back to a metastate state, $A$, starting with a randomly chosen state
with overlap $q$ with state $A$}. The data shown here are the results
of a direct numerical simulations of a finite size arrays with $\alpha = 1.0$; 
the gray, dashed and black curves correspond to $N = 7, 17$ and $23$
respectively.} 
\vspace{0.5in} 

\section{The Physical Consequences}

What are the observable consequences of the different behavior
in the response and the correlation functions for the
three cooling regimes?  If the periodic array is placed
in a time-dependent field $H(t)$ that is small
compared to its constant part, the resulting Josephson
currents generate a total magnetic moment
\beq
M_t = \mu \int_{\infty}^{t} g_{tt'} d_{tt'} \left\{H(t) - H(t') \right\} dt'
\label{moment}
\eeq
which can be measured with an inductance coil; here
\beq
\mu = \left(\frac{2e}{\hbar e}\right)^2
\left(\frac{L^2}{12}\right)^2
N
\left(\frac{J_0^2}{T}\right)
\frac{1}{\alpha^2}
\left( 1 -\frac{J_0^2}{A^2 \alpha}\right)
\left( \frac{t_0}{\alpha^{1/2}\theta_t}\right)
\label{mu}
\eeq
where $L = N l$.
The form
of $M$ in (\ref{moment}) should be contrasted with the response
of the spin systemto a physical magnetic field which is just
$g_{tt'}$; the dissimilarity is due to the fact that in the
Josephson array {\sl direct} coupling of the physical field
to the ``spins'', $s_i = e^{i\phi}$, is prohibited
by gauge invariance.
Instead $H$ couples to the network via changes of the Josephson
exchanges;
hence the response involves a four-spin term ($GD$).
Here we discuss the magnetization induced by a small increment
of magnetic field applied at time $t_H$; this is of
course the simplest possible case.  Still the value of
the observed moment will depend on the
relative magnitudes of $t_H$, $t_c$, $t^*$ and $t$ as indicated
in the schematic of Figure 16.  We proceed to discuss each separate
case below.

\centerline{\epsfxsize=8cm \epsfbox{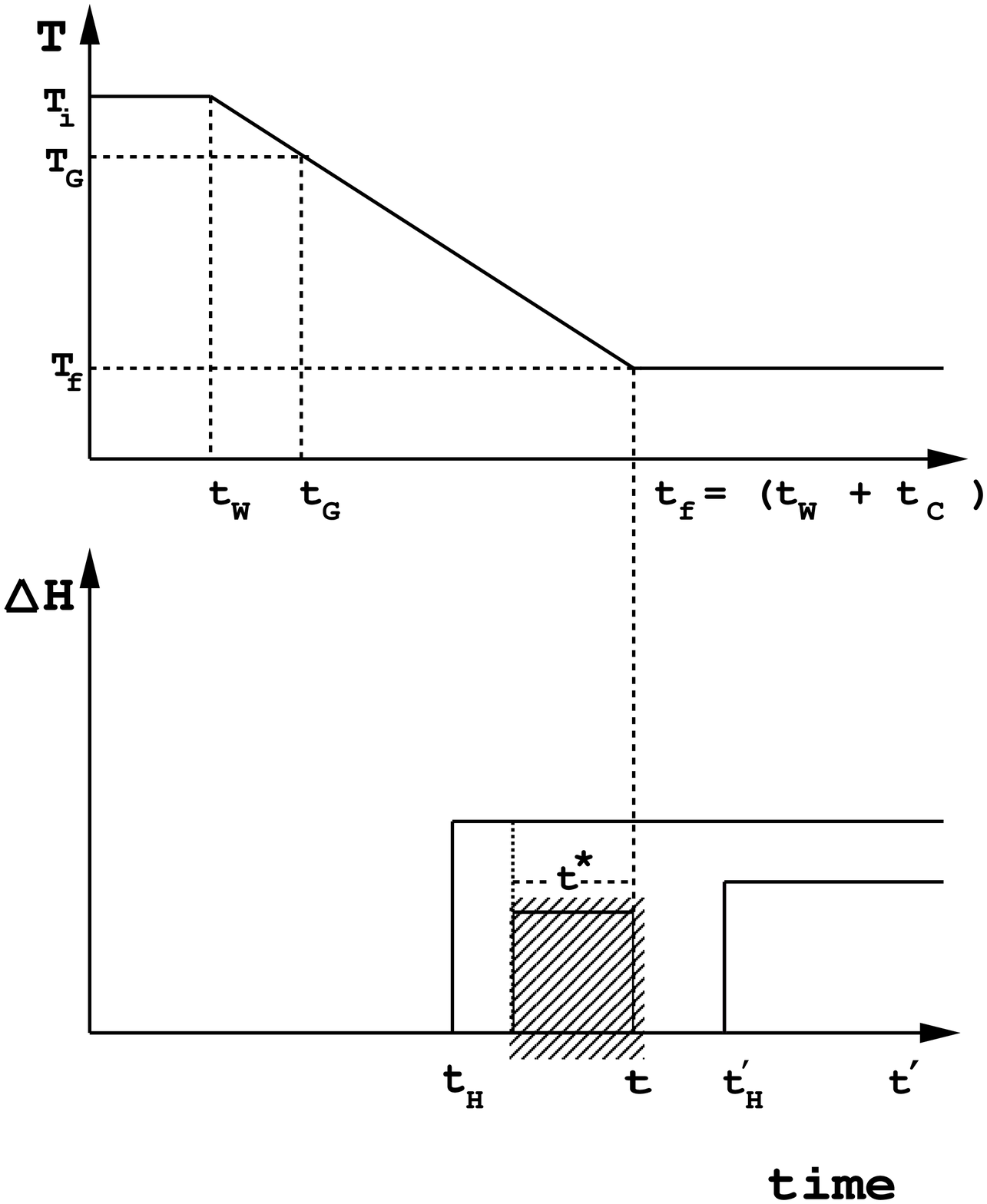}} 
{\footnotesize {\bf Fig 16. Schematics of (a) Temperature ($T$)
(b) Application of an Additional Field   ($\Delta H$) vs time}
indicating the time-scales
involved in a finite-cooling experiment as discussed in
the text; $T_G$ and $t_G$ refer to the 
temperature and time when the system goes out of equilibrium, 
and $t_H$ and $t^*$
are the time-scales associated with the onset of an additional
field and ageing; t is the total time associated with a given
measurement.
}\vspace{0.5in} 

\subsection{$t, t_H \gg t_c$}

Here we apply a field after a fast quench, 
a regime where there is a generalized fluctuation dissipation relation.
Then the expression for the moment in (\ref{moment}) becomes
\beq
M = \mu x \int_{t_c \sim 0}^{t_H} \frac{1}{2} (D^2)' \Delta H dt' = 
\frac{\mu}{2} x d^2_0 \left(\frac{t_H - t_G}{t - t_G}\right)^{2\gamma}
\Theta(t_H- t_G)\Delta H
\label{fmoment} 
\eeq
where we have used the ``fast cooling'' scaling form for the correlation
function exhibited in Table I.  Since the fast-cooling response function
decays with time, (\ref{fmoment}) indicates an ``ageing'' moment
which decreases as a function
of decreasing ratio $\frac{t_H - t_G}{t - t_G}$ ($t_H,t > t_G$).
(\ref{fmoment}) also implies 
markedly different behavior if $t_H > t_G$  or $t_H < t_G$;
this is the array analogue of the standard zero-field vs. field 
cooled susceptibility measurement often performed on 
spin glasses.\cite{Fischer91}

\subsection{$t_H < t_c = t$}

Now the cooling is continued after the field application (cf. Figure 13),
and the dominant contribution to the expression for the moment, (\ref{moment}),
will come from the range $t_c - t^* < t' < t_c$ where $t^*$ is
the ``ageing'' time-scale discussed in the slow cooling regime in Section III.
Here there are two possibilities:

{\sl (a) $t_c - t_H  < t^*$}

In this case the correlation and the response functions are non-decaying
(cf. Figs. 4-7) and the moment is a constant as in (\ref{mconst}).

{\sl (b) $t_c - t_H > t^*$}

The correlation and response functions have no memory on time-scales
greater than $t^*$; therefore the moment is zero.  The system
is displaying ageing analogous to that observed in spin
glasses;\cite{Fischer91} 
as in the latter case, there appears to be a ``rejuvenation'' associated
with improved memory at low temperatures where $t^*$ increases.  

We also remark that application of a a.c. field
could lead to a direct experimental determination of $G_\omega D_\omega$; 
there we expect $\chi''(\omega) \sim f(\omega t^*)$
as shown below in Figure 17. 
Here we define $\chi(\omega) = \int_{-\infty}^{t} dt' \exp^{i\omega (t
- t')} d_{tt'} g_{tt'}$.

We note that the low-frequency part of
the displayed curves is universal and so would provide an experimental
determination of the time-scale associated with ageing, $t^*$.

\centerline{\epsfxsize=8cm \epsfbox{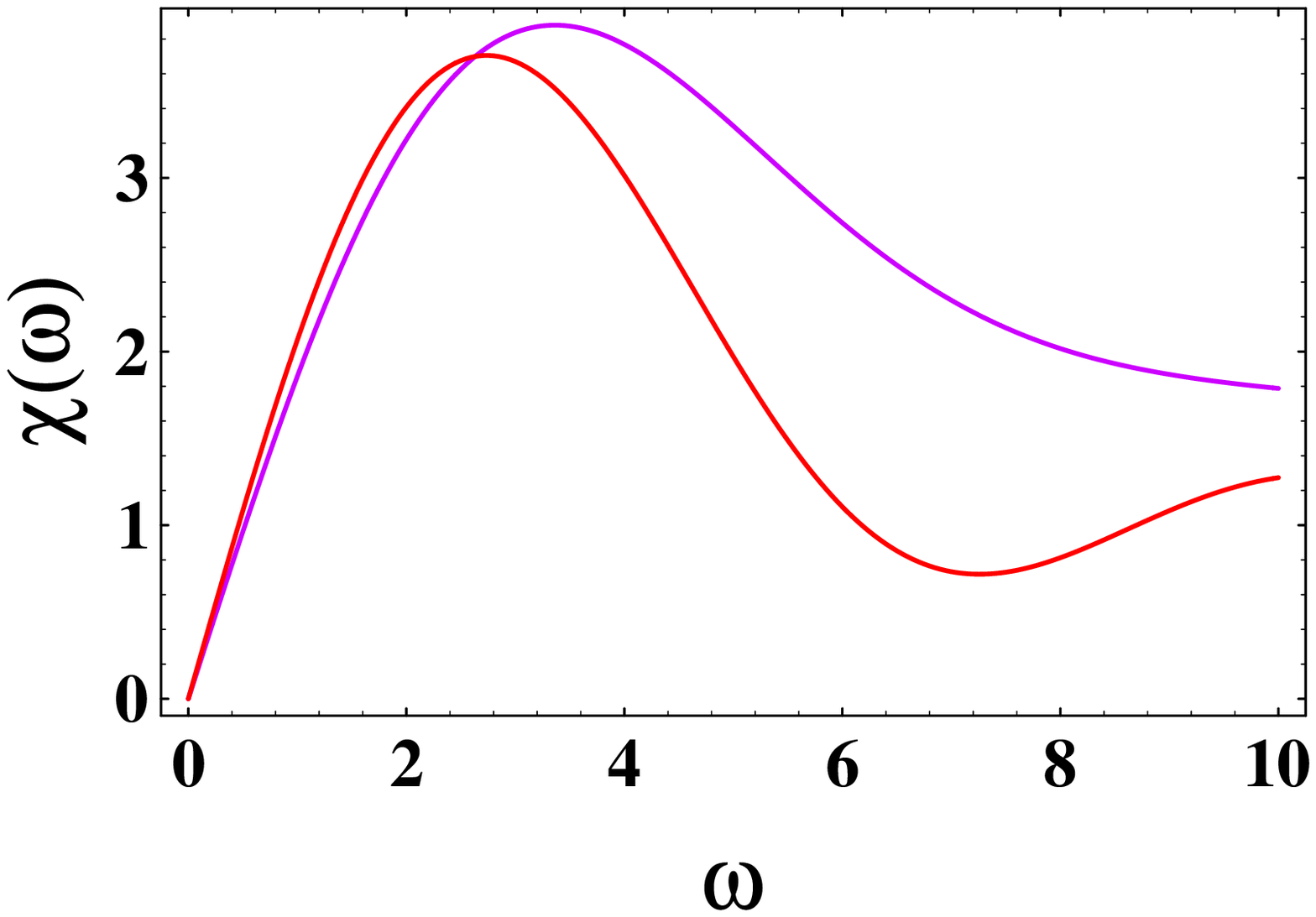}} 
{\footnotesize {\bf Fig 17. Typical imaginary parts of ac susceptibility for
the slow cooling processes}. Here we show the results for the cooling rates 
$\Gamma=0.00057$ (upper curve) and $\Gamma=0.0022$ (lower curve). The upper 
curve shows imaginary part of susceptibility measured at reduced temperature
$\theta_f=1.0$; the lower curve corresponds to the measurement at
$\theta_f=1.2$. The frequency $\omega$ is measured in units of the inverse
characteristic time, 
$t^*=120$, corresponding to the decay of the correlation
function. Here we define it by the following procedure: (i)
we do a linear fit to $d_{tt'}$; (ii) we identify
the intersection of this approximated version of $d_{tt'}$
and the x-axis as $t - t^*$.
Using this definition,
$t^*$ acquired the identical value for the two cooling processes
considered here.
}\vspace{0.5in}

\subsection{$t > t_c > t_H$}

In this ``crossover regime'' the incremental time-dependent field is applied
before the system has reached its final cooling temperature ($T_f < T_G$);
then the expression for the moment becomes
\beq
M \sim \int_{t_G}^{t_H} \tilde{\Gamma} G[\tilde{\Gamma}(t - t')] 
D[\tilde{\Gamma}(t - t')] \Delta H dt' =  \mu_A \Delta H
\label{mconst}
\eeq
where $\mu_A$ is a number; this result is obtained by changing the integration
variable to $\tilde{t} = \tilde{\Gamma} t'$ where 
$\tilde{\Gamma} \sim \left(\frac{d\tau}{dt}\right)^{1 - \eta(T_f)}$. In this
regime the moment is simply a constant with no time-dependence but
with possible dependence on the cooling rate if $\eta \neq 0$.

\section{Concluding Discussion}

In conclusion we have studied history-dependence and ageing
in a periodic model in the {\sl absence} of disorder.  In the
case of a fast quench ($ t \gg t_c$) a  generalized fluctuation
dissipation theorem relation ($G \sim x D'$) exists between
the correlation and the response functions, and furthermore
the latter exhibit a predicted scaling form.\cite{Cugliandolo93}
In the other limit of slow cooling ($t = t_c$) there is numerical
evidence for the existence of an ``ageing'' time-scale $t^*$
such that for $t > t^*$ the system forgets its past, but for $t < t^*$
it has ``perfect memory''.
This additional time-scale
increases to that associated with the cooling process with
decreasing temperature, 
indicating that at low temperatures the system
recovers its memory and an adiabatic solution (cf. Table I)
reappears. However the form of this solution
does not ressemble (cf. Fig. 8) the ``adiabatic ansatz''\cite{Feigelman95}  
$G \sim \delta$
and $D \sim D_0$, which is probably {\sl never} correct.
We speculate that this
is due to the presence of an extensive number of
states 
exactly
at the glass transition,
where the time-scale associated with the cooling process is
significantly less than that of the system's relaxation.  By contrast,
in the Sherrington-Kirkpatrick model such ``branching of states''
occurs at lower
temperatures where the cooling and relaxation times are comparable;
in this case the assumption that there are no additional
time-scales\cite{Ioffe88} leads to results which agree with those
obtained by thermodynamic methods.\cite{Mezard87}

We also speculate on two different scenarios for the decrease
of the exponent $\eta$ ($t^* = \left(\frac{dt}{d\tau}\right)^{1 - \eta(T)}$)
with decreasing temperatures, and the reappearance of the adiabatic
solution.   One is that the latter solution is always present for
$\theta >
\theta_G$, but its weight is proportional to a high power of $\theta - \theta_G$,
and thus is only observable at lower temperatures.  Another possibility
is that $\eta \rightarrow 0$ corresponds to a second dynamical
transition at lower temperatures.
Below
the temperature associated with this second dynamical instability the
metastable states would not evolve dramatically with subsequent cooling,
leading to the absence of further ageing effects. Naturally the emergence
of $t^*$ from an analytic ansatz for the dynamical equations would
provide additional insight to this issue.  Unfortunately these
equations are difficult to solve in the slow-cooling regime since
the corrections to long-time behavior are of the same order as
the leading terms themselves.

Just below its glass transition temperature, the periodic array
becomes stuck in one of an extensive number of metastable states that
are equidistant in phase space.  Our numerical studies indicate
that there is {\sl no further subdivision} of these states at lower
temperatures, in contrast to the situation for the (disordered)
Sherrington-Kirkpatrick model.  However the probability to evolve
to different metastable states remains finite even if the starting
overlap is very close to unity, which suggests that the distribution
of barriers and the overlaps are {\sl not} simply related.
It would be interesting to know if these features of 
the periodic array's low-temperature phase space are signatory
of a glass with distinct dynamic and static transitions.
 
Naturally there remain several questions to be addressed in
future projects.  A key feature of this periodic array is
that it can be realized experimentally;\cite{Chandra96b} a
calculation and subsequent experimental determination of
its nonlinear response be quite interesting.  In a real
system there are everpresent induced screening currents
ignored in the present treatment, and the addition
of these terms could lead to novel memory effects. Amusingly
enough, ageing effects are not present
in the disordered long-range array; a study of the crossover between
the periodic and the random networks might
provide further insight into the origin of this widespread
but mysterious phenomenon.

We would like to acknowledge useful discussions with J. Kurchan,
D. Sherrington,
M. Stephens, and thank A. Schweitzer for his help in optimizing
the numerical code. 
Financial support via the grants from RFBR \# 95-02-05720 and
INTAS-RFBR \# 95-0302 (M.V.F. and D.M.K.) and via DGA grant \#
94-1189 (M.V.F.) is gratefully acknowledged.
M.V. Feigelman and D.M. Kagan thank
NEC for hospitality.

\end{document}